\documentclass[oldversion]{aa}

\usepackage{graphicx}
\usepackage{natbib}
\usepackage{amssymb}
\usepackage{amsmath}
\usepackage{rotating}
\usepackage[dvips=true, bookmarks=true, bookmarksnumbered=true]{hyperref}
\usepackage{txfonts}
\usepackage{fixltx2e}

\newenvironment{narrow}[2]{%
	\begin{list}{}{%
	\setlength{\topsep}{0pt}%
	\setlength{\abovecaptionskip}{5pt}%
	\setlength{\leftmargin}{#1}%
	\setlength{\rightmargin}{#2}%
	\setlength{\listparindent}{\parindent}%
	\setlength{\itemindent}{\parindent}%
	\setlength{\parsep}{\parskip}}%
	\item[]}
	{\end{list}
	}

\begin{document}

\title{Line formation in solar granulation}

\subtitle{VII. CO lines and the solar C and O isotopic abundances}

\author{Patrick C. Scott\inst{1}
\and
Martin Asplund\inst{1}
\and
Nicolas Grevesse\inst{2}$^,$\inst{3}
\and
A.~Jacques Sauval\inst{4}}

\institute{Research School of Astronomy and Astrophysics, Mt. Stromlo Observatory, Cotter Rd., Weston Creek, ACT 2611, Australia
\and
Centre Spatial de Li{\`e}ge, Universit{\'e} de Li{\`e}ge, avenue Pr{\'e} Aily, B-4031 Angleur-Li{\`e}ge, Belgium
\and
Institut d'Astrophysique et de G{\'e}ophysique, Universit{\'e} de Li{\`e}ge, All{\'e}e du 6 ao{\^u}t, 17, B5C, B-4000 Li{\`e}ge, Belgium
\and
Observatoire Royal de Belgique, avenue circulaire, 3, B-1180 Bruxelles, Belgium
}

\date{Received 9 February 2006 / Accepted 18 April 2006}

\abstract{
CO spectral line formation in the Sun has long been a source of consternation for solar physicists, as have the elemental abundances it seems to imply.  We modelled solar CO line formation using a realistic, ab initio, time-dependent 3D radiative-hydrodynamic model atmosphere.  Results were compared with space-based observations from the ATMOS space shuttle experiment.  We employed weak $^{12}$C$^{16}$O, $^{13}$C$^{16}$O and $^{12}$C$^{18}$O lines from the fundamental (\mbox{$\Delta v = 1$}) and first overtone (\mbox{$\Delta v = 2$}) bands to determine the solar carbon abundance, as well as the $^{12}$C/$^{13}$C and $^{16}$O/$^{18}$O isotopic ratios.  A weighted solar carbon abundance of \mbox{$\log\epsilon_\mathrm{C}=8.39\pm 0.05$} was found.  We note with satisfaction that the derived abundance is identical to our recent 3D determination based on C\,\textsc{i}, [C\,\textsc{i}], C$_2$ and CH lines, increasing our confidence in the accuracy of both results.  Identical calculations were carried out using 1D models, but only the 3D model was able to produce abundance agreement between different CO lines and the other atomic and molecular diagnostics.  Solar $^{12}$C/$^{13}$C and $^{16}$O/$^{18}$O ratios were measured as 86.8$^{+3.9}_{-3.7}$ (\mbox{$\delta^{13}\mathrm{C} = 30^{+46}_{-44}$}) and 479$^{+29}_{-28}$ (\mbox{$\delta^{18}\mathrm{O} = 41^{+67}_{-59}$}), respectively.  These values may require current theories of solar system formation, such as the CO self-shielding hypothesis, to be revised.  Excellent agreement was seen between observed and predicted weak CO line shapes, without invoking micro- or macroturbulence.  Agreement breaks down for the strongest CO lines however, which are formed in very high atmospheric layers.  Whilst the line asymmetries (bisectors) were reasonably well reproduced, line strengths predicted on the basis of C and O abundances from other diagnostics were weaker than observed.  The simplest explanation is that temperatures are overestimated in the highest layers of the 3D simulation. Thus, our analysis supports the presence of a COmosphere above the traditional photospheric temperature minimum, with an average temperature of less than 4000K.  The shortcoming of the model atmosphere is not surprising, given that it was never intended to properly describe such high layers.
}

\titlerunning{Solar line formation: VII.  CO lines and isotopic ratios}
\authorrunning{P. Scott et al.}

\offprints{martin@mso.anu.edu.au}

\maketitle
%

\section{Introduction}
\label{sec1}

\subsection{CO in the Solar Spectrum}
\label{preintro}
Thousands of vibration-rotation lines from the fundamental $^1\Sigma^+$ electronic band of CO are present in the infrared solar spectrum obtained from space in the Atmospheric Trace Molecule Spectroscopy (ATMOS) space shuttle mission (see Sects.~\ref{spectral} and \ref{atmos}). From 1350 to \mbox{2328 cm$^{-1}$} (i.e.~4.3 to \mbox{7.4 $\mu$m}; fundamental bands from 1--0 to 20--19) and 3410 to \mbox{4360 cm$^{-1}$} (i.e.~2.3 to \mbox{2.9 $\mu$m}; first overtone bands from 2--0 to 14--12), the solar spectrum looks more like a pure CO absorption spectrum, displaying not only $^{12}$C$^{16}$O lines but also $^{13}$C$^{16}$O, $^{12}$C$^{18}$O and $^{12}$C$^{17}$O isotopomeric lines.  Thanks to their very high excitations (higher than in any laboratory spectra where rotational excitations are concerned), the solar CO lines have been used to derive a new set of highly accurate molecular constants \citep{farrenq91, sauval92}.  As the most sensitive indicator of solar photospheric temperatures, and formed over an extremely wide domain of optical depths (Fig.~\ref{formationheights}), analysis of the CO lines offers a unique opportunity to test physical conditions up to very high atmospheric layers, derive isotopic abundance ratios and refine the solar carbon abundance \citep{GandS91, GandS92, GandS94, GandS95}.  Only a limited number of clean CO lines are available in ground-based spectra however, with the $\Delta v = 1$ sequence heavily masked by strong, broad telluric absorption and the $\Delta v = 2$ sequence also strongly polluted by terrestrial lines.

\subsection{CO Analyses and Problems}
\label{intro}
Since the seminal observations of \citet{NH} revealed limb CO core brightness temperatures below the classical photospheric temperature minimum of 4500K, debate has raged over what CO absorption features in the solar spectrum really tell us.  \citet{AT81} found that the most natural explanation was an absence of any photospheric temperature minimum, requiring an outwardly decreasing temperature structure from the base of the photosphere right through the chromosphere.  However, given that Ca\,\textsc{ii} and Mg\,\textsc{ii} line cores indicate chromospheric temperatures well above the classical minimum \citep{AL76}, \citet{Ayr81} proposed the existence of a cool CO structure in the low chromosphere, coexistent with other areas of hotter gas responsible for the Ca\,\textsc{ii} and Mg\,\textsc{ii} core emission: a `thermally bifurcated' low chromosphere.

\citet{AW89} went on to test the LTE assumption of the bifurcation model, finding that non-LTE effects had $<$2$\%$ effect upon CO line cores, even as near to the solar limb as $\mu = 0.1$.  Whilst these findings were disputed \citep{MAL90}, the existence of some cool CO in the atmosphere was accepted, though temperature inhomogeneities were suggested to be due to the effects of localised mechanical heating rather than CO radiative cooling.  The presence of cool CO in the chromosphere was put beyond doubt four years later with the discovery of off-limb CO emission indicating temperatures of less than 4000K in some places \citep{SLA}.

Using imaging spectroscopy, \citet{UNR} studied spatial and temporal variations in CO absorption at disk centre, observing temperature inhomogeneities of up to 600K correlated with hydrodynamic perturbations.  This result supported the existence of the temperature bifurcation, though driven more by dynamic processes than CO cooling and magnetic heating as suggested originally.  This was picked up on in a subsequent revision of the bifurcation model \citep{AR96} in which the cool component was dubbed the `COmosphere'.

\citet{CS95, CS97} demonstrated that chromospheric Ca\,\textsc{ii} H$_{2V}$ and K$_{2V}$ grains are formed by acoustic shocks, using a 1D non-LTE hydrodynamic model.  Their results strongly supported the notion pushed by \citet{AR96} of a bifurcated atmosphere with differences caused by dynamic effects, even also suggesting some influence by CO cooling owing to the inclusion of instantaneous chemical equilibrium (ICE) CO opacities in the model.  \citet{Kalk99} and \citet{Kalk2001} argued against the Carlsson-Stein model because it did not adequately represent the entire solar energy output, and so could not reproduce the observations of ubiquitous chromospheric emission.  However, this was never \citeauthor{CS97}'s intention, as discussed by \citet{Ayr2002} in an extensive rebuttal of \citeauthor{Kalk99}'s criticism.  \citeauthor{Ayr2002} went on to show that despite not being designed with CO diagnostics in mind, the Carlsson-Stein model far outperformed \citeauthor{Kalk99}'s preferred model \citep[VAL;][]{VAL73, VAL76, VAL81} when put to work on solar CO lines.

Meanwhile, \citet{UitII} performed detailed 1D NLTE simulations of CO absorption in the solar disk, reconfirming that the LTE approximation is appropriate in this context.  \citet{UitI} also studied 3D LTE CO line formation using a single snapshot from an early version \citep{SandN89} of the current state-of-the-art 3D atmospheric models \citep{SandN, AspI}.  The calculated line profiles were deeper than those observed by the ATMOS mission, something \citeauthor{UitI} attributed to the ICE approximation.

Growing concern \citep{UitI, UitII, Ayr2002} over the appropriateness of the ICE approximation was addressed by \citet{AR2003}, who investigated departures from ICE in CO lines using the Carlsson-Stein model.  Results showed significant concentrations of cool CO gas extending just \mbox{700 km} above the Sun's surface, with the ICE assumption valid below this height.  2D simulations with non-equilibrium chemistry \citep{2D} have confirmed that the ICE approximation is valid in the photosphere and low chromosphere.  Despite employing non-equilibrium chemistry in both the hydrodynamic and radiative transfer simulations, modelled CO lines were still considerably deeper than observed profiles. This is probably because the temperature structure of the excerpt from their 2D model is too cool, producing an effective surface temperature more than 300K lower than the true solar value.  Furthermore, 2D simulations tend to give erroneous photospheric line profiles compared with 3D calculations because of the geometric restriction on the convective flow \citep{AspRes}.

\subsection{Solar C and O Isotopic Abundances}
\label{introabuns}
The solar carbon abundance has recently been revised \citep{AspRev} from the \mbox{$\log\epsilon_\mathrm{C} = 8.52 \pm 0.06$} given by \citet{GandS98}.  Analyses of [C\,\textsc{i}], C\,\textsc{i}, CH and C$_2$ \citep{CtoO, AspVI} now indicate that the logarithmic solar carbon abundance is in fact $8.39 \pm 0.05$.  These results provide a comparison for the solar carbon abundance determination we carry out using CO lines.  Considering the difficulty of reproducing observed CO lines with theoretical calculations, these provide a very stringent test of the model atmosphere.  Asymmetries of the CO lines constitute an even greater challenge for theoretical atmospheres \citep{GSbi}.

Early measurements of the solar $^{12}$C/$^{13}$C ratio using CH lines were typically higher than the terrestrial reference ratio of $89.4 \pm 0.2$ \citep{IUPAC02}, and exhibited large uncertainties (see \citealt{HLG87} for an inventory of early measures).  The most reliable (lowest uncertainty) determinations to date utilised CO lines, producing near-terrestrial ratios of $84 \pm 9$ \citep{Hall73} and $84 \pm 5$ \citep{HLG87}.  The solar $^{16}$O/$^{18}$O ratio was also measured, producing values of $>500$ and $440 \pm 50$ respectively.  These are similar to the representative terrestrial ratio of $498.7 \pm 0.1$ \citep{IUPAC02}.\footnote{The geological manner of expressing such measurements is in `permil', a deviation in parts per thousand from the terrestrial value:
\begin{equation}
\label{permil}
\delta^b\mathrm{X} = 1000 \times \left( \frac {(^b\mathrm{X}/^a\mathrm{X})_\mathrm{sample}} {(^b\mathrm{X}/^a\mathrm{X})_\oplus} - 1\right)
\end{equation}
where X is the element in question, $a$ refers to its reference isotope and $b$ to the isotope in question.  It should be realised that the ``terrestrial ratios'' given above are from standard substances chosen to define the zero-point of this scale; in reality, the terrestrial isotopic composition varies far more than the errors attached to these values appear to suggest \citep{IUPAC98, IUPAC02}.}

Different lunar regolith (surface) analyses of the solar wind have indicated $\delta^{13}\mathrm{C}\leq -105\pm20$ \citep{hash04} and $30\leq\delta^{13}\mathrm{C}\leq-30$ \citep[combined results reviewed in][]{WBBW}.  $\delta^{18}$O values tabulated in the same review for direct measurements of solar wind particles by two different satellites give 110$^{+450}_{-250}$ and 120$^{+280}_{-190}$.  For comparison, the results of \cite{HLG87} on the same scale imply \mbox{$\delta^{13}\mathrm{C} = 59^{+67}_{-60}$} and \mbox{$\delta^{18}\mathrm{O} = 130^{+145}_{-115}$}, without the inherent uncertainty in the scale taken into account.  The most recent lunar regolith analyses differ in their estimates of solar $\delta^{18}$O, with one suggesting a negative value (\citealt{HC05}, see also \citealt{Davis}) and the other producing $\delta^{18}\mathrm{O} = 50$ \citep{Ireland06}.  Fractionation that might occur in the solar wind is still not properly quantified however \citep{WBBW}, so such results may say as much about this process as they do about the actual photospheric ratios.  Solar chemical evolution is also thought to effect abundance agreement between the current photosphere and lunar inclusions irradiated by the prehistoric solar wind, though the predicted difference over the Sun's lifetime is less than 10 permil in $\delta^{13}$C or $\delta^{18}$O \citep{turcotte02}.  Recently, \citeauthor{YK04} (\citeyear{YK04}, see also \citeauthor{Yin} \citeyear{Yin}) proposed a model of solar system formation whereby oxygen isotopic fractionation occurs in the protosolar cloud due to UV irradiation.  This theory predicts a solar $\delta^{18}$O of $-$50, something that can be directly tested with the $^{18}$O/$^{16}$O measurement to be performed herein.

\subsection{This Study}
\label{introthispaper}
In this paper, we investigate CO absorption line formation in the solar spectrum using a three-dimensional radiation-hydrodynamic simulation of the Sun's atmosphere.  This analysis is similar to that of \citet{UitI}, but with an improved 3D model as its basis.  This enables a better description of CO and a more precise evaluation of the reality of the model in the context of CO, as well as simulation of line formation in the disputed COmosphere.  It should be noted that the model is not intended to reproduce the ubiquitous chromospheric emission nor near-limb CO lines, as it does not include magnetic fields nor extend to great enough heights.

Using the same model and data, we also arrive at new estimates of the solar carbon abundance, $^{12}$C/$^{13}$C and $^{16}$C/$^{18}$O ratios using appropriate sets of weak $^{12}$C$^{16}$O, $^{13}$C$^{16}$O and $^{12}$C$^{18}$O lines.  Unfortunately, the $^{12}$C$^{17}$O lines clearly identified in the ATMOS spectra are too weak and perturbed to permit any defensible estimate of the solar $^{16}$O/$^{17}$O ratio, in our opinion.  The isotopic ratios constitute an improvement over the 1D analyses of \citet{Hall73} and \citet{HLG87}, thanks to the use of 3D model atmospheres, more recent $gf$ values, more appropriate line lists and better observations.  We believe that our ratios are also superior to alternative determinations carried out very recently by \citet{Ayres05}, as will be detailed in Sect.~\ref{comparisondiscuss}.  The determination of the solar carbon abundance using CO lines provides an important comparison with the study of \citet{AspVI} based on [C\,\textsc{i}], C\,\textsc{i}, CH and C$_2$ lines.  It also circumvents past trouble in using CO lines and 1D models to measure the solar carbon abundance \citep[e.g.][]{GandS95}.

The 3D model and spectral line lists analysed are briefly described in Sect.~\ref{model} and Sect.~\ref{spectral} respectively, whilst Sect.~\ref{atmos} details our manipulation of the ATMOS data.  The study of the detailed shapes of strong CO absorption lines and their implications for the temperature structure in very high atmospheric layers is presented in Sect.~\ref{coresults}. In Sect.~\ref{abunresults} we present abundance and isotopic ratio determinations, based on weak CO lines.  Comparisons with earlier work are given in Sect.~\ref{comparisondiscuss}, and our conclusions are summarised in Sect.~\ref{conclusions}.  A detailed description of the apodization procedure applied to modelled spectra is given in Appendix~\ref{ATMOSappendix}.  Appendix~\ref{scalefactors} provides derivations of scaling factors used in the analysis, and Appendix~\ref{lists} contains full line lists.

\section{Model Atmospheres and Line Formation Calculations}
\label{model}
The hydrodynamic simulation used was described by \citet{AspI}.  Specifically, it covers a physical area \mbox{6 $\times$ 6 $\times$ 3.8 Mm} of which about \mbox{1 Mm} is above the optical solar surface ($\tau_\mathrm{Ross} \approx 1$), at a resolution of \mbox{200 $\times$ 200 $\times$ 82}.  The simulated domain was bounded below by a transmitting boundary, and above by an extended transmitting boundary across which the density gradient was kept hydrostatic.  Horizontal boundaries were periodic.  The MHD equation-of-state \citep{MHDI, MHDII} and Uppsala opacities were used, with LTE assumed.  Continuous and line opacities were calculated using opacity binning.  The model included no tuneable free parameters, and was characterised as solar by the accepted gravity, effective temperature and standard solar composition of \citet{GandS98}.  A numerical viscosity was employed to stabilise the simulation.  About 100 snapshots of the convective simulation were stored, representing approximately \mbox{50 min} of real solar time.  Radiative transfer calculations for spectral line formation were carried out over an interpolated \mbox{50 $\times$ 50 $\times$ 82} grid, using the Uppsala equation-of-state and opacities with molecular densities determined under the ICE approximation.  LTE was again assumed, as validified for the CO lines by \citet{AW89} and \citet{UitI}.  This is the same atmosphere and line formation code recently used to re-derive the solar abundances of all elements between Li and Ca (e.g. \citealt{AspIII, AspV}; \citealt{AspII, AspIV, AspRev, AspVI}; Asplund et al., in preparation; \citealt{APForbidO, CtoO}).

For comparative purposes, abundance calculations were carried out using four different model atmospheres: the 3D hydrodynamic model, the 1D HM model \citep{HM}, the 1D \textsc{marcs} model \citep{MARCS75, MARCS97}, and a contraction of the 3D model into the vertical dimension only, which we designate `1DAV'\footnote{`One Dimensional AVerage'}.  The horizontal averaging used to produce the 1DAV model was performed over surfaces of common optical depth rather than geometrical height, as the optical depth scale has most relevance to line formation.  All three 1D models included microturbulence of \mbox{1 km\,s$^{-1}$}.  HM and \textsc{marcs} were largely insensitive to microturbulence, with abundances reduced by only 0.01--0.02 dex going from $\xi_\mathrm{t}=0$ to $\xi_\mathrm{t}=1~\mathrm{km~s}^{-1}$, whilst the 1DAV model exhibited an increase of 0.06--0.07 dex in abundance with $\xi_\mathrm{t}=1~\mathrm{km~s}^{-1}$ instead of $\xi_\mathrm{t}=0$.  Microturbulence was not fitted for in the 1D models, as they were used simply for comparative purposes rather than highly accurate determinations. As it turned out, the model most sensitive to microturbulence (1DAV) exhibited the least trends in equivalent width and excitation potential anyway.  The oxygen and nitrogen abundances used for the 3D, HM and \textsc{marcs} models were indicative of the most recent values produced with each model \citep[][where in the case of oxygen, values are from vibration-rotation and pure rotational OH lines only]{AspRev}.  The oxygen abundance used for the 1DAV model was derived self-consistently from the same OH lines used to set the oxygen abundances of the other models, and employed previously by \citet{AspIV}.  The nitrogen abundance for the 1DAV model was simply approximated as the 3D value in the absence of any appropriate line calculations.  Oxygen and nitrogen abundances used in the line calculations with the different models are shown in Table~\ref{OandN}.

\begin{table}[tbp]
\centering
\caption[O and N abundances used for isotopic determinations]{Oxygen and nitrogen abundances used in line formation calculations for each model in this study.  3D, HM and \textsc{marcs} values are from \citet{AspIV}, 1DAV oxygen calculated using the same lines as \citet{AspIV} and 1DAV nitrogen abundance simply set to the 3D value.}
\label{OandN}
\begin{tabular}{l c c c c}
\hline
Model & 3D & HM & \textsc{marcs} & 1DAV\\
\hline
$\log \epsilon_\mathrm{O}$ & 8.66 & 8.85 & 8.80 & 8.70 \vspace{0.5mm}\\
$\log \epsilon_\mathrm{N}$ & 7.80 & 7.95 & 7.90 & 7.80 \vspace{0.5mm}\\
\hline
\end{tabular}
\end{table}

\section{Spectral Lines}
\label{spectral}

\begin{figure}[tbp]
\centering
\includegraphics[width = 0.45\textwidth]{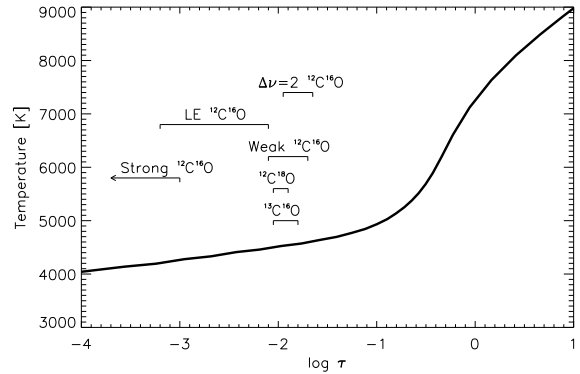}
\caption[Relative formation heights of line lists used]{Approximate optical depths of core formation for the different CO line lists used in Sect.~\ref{coresults} (strong $^{12}$C$^{16}$O) and Sect.~\protect\ref{abunresults} (all others).  The temperature structure shown is that of the standard 3D model atmosphere \protect\citep{AspI}.}
\label{formationheights}
\end{figure}

We utilise six distinct CO line lists, with $gf$ values calculated from \citet{GChack94}.  We adopted the $^{12}$C$^{16}$O dissociation energy of \citet[][\mbox{11.108 eV}]{D087}.  All lines were selected to be free of blends.  Typical formation heights (estimated for the \textsc{marcs} model atmosphere) are shown in Fig.~\ref{formationheights}.  CO line formation in higher atmospheric layers is examined in Sect.~\ref{coresults} via bisector analysis of 31 $^{12}$C$^{16}$O lines (Table~\ref{biseclines}).  These are all strong lines that are formed high in the atmosphere.  As the height of their formation and hence velocity signatures strongly depend on the temperature structure of the atmosphere, these lines can probe the disputed `COmosphere' region.

Abundance calculations in Sect.~\ref{abunresults} draw upon the five remaining lists, with equivalent widths measured in the ATMOS ATLAS-3 spectrum (cf. Sect.~\ref{atmos}).  The primary investigation is performed with a set of 13 weak $^{12}$C$^{16}$O lines (Table~\ref{weaklines}) of high excitation, as well as sets of 16 $^{13}$C$^{16}$O (Table~\ref{1316}) and 15 $^{12}$C$^{18}$O (Table~\ref{1218}) lines.  These lines all have significantly lower formation heights than the strong $^{12}$C$^{16}$O lines considered for Sect.~\ref{coresults} (Fig.~\ref{formationheights}).  Being CO features, in absolute terms they still form reasonably high in the atmosphere, but were selected as forming low enough for abundance determination using the 3D model.

The further two lists consist of 15 low excitation (LE) $^{12}$C$^{16}$O (Table~\ref{LElines}) and 66 first overtone (\mbox{$\Delta v = 2$}) $^{12}$C$^{16}$O (Table~\ref{dv2lines}) lines.  The LE lines are formed across a broad optical depth range high in the atmosphere (due to the low temperatures required for significant populations in the lower energy levels), and the \mbox{$\Delta v = 2$} at around the same height as the three primary sets (Fig.~\ref{formationheights}).  These supplementary lists are used to derive alternative carbon abundances (and hence isotopic ratios) to the primary (weak) $^{12}$C$^{16}$O list, exploring the abundance performance of the model over higher layers and overtone bands.  The three different $^{12}$C$^{16}$O line lists, which also have different temperature sensitivities, have traditionally produced widely varying carbon abundance measures using 1D models, so a consistent result in 3D would greatly increase confidence in the new C abundance and isotopic ratios.

\section{The ATMOS Infrared Solar Spectrum}
\label{atmos}

The ATMOS instrument is a Fourier transform spectrograph (FTS).  It was carried on a series of space shuttle missions between 1985 and 1994, retrieving pure solar and atmospheric occultation spectra between 625 and \mbox{4800 cm$^{-1}$}.  Given the instrumental resolution (\mbox{0.01 cm$^{-1}$}, \citealt{ATMOS94}) and taking a representative wavenumber of \mbox{2000 cm$^{-1}$} (corresponding to a wavelength of \mbox{5000 nm}) for the section of the spectrum predominantly considered in this paper, the instrument's resolving power for our purposes is:
\begin{displaymath}
R = \frac{k}{\Delta k} = \frac{2000}{0.01} = 200\,000.
\end{displaymath}

The solar disk-centre spectrum from the 1994 ATMOS ATLAS-3 mission\footnote{\protect\url{http://thunder.jpl.nasa.gov/atmos/at3.solar}} was normalised, with the continuum level defined as the highest intensity point in the extracted section.  Given the high signal-to-noise of the data ($\sim$1000:1 around \mbox{5000 nm} by our measurements), detailed averaging was deemed unnecessary.  The Sun's gravitational redshift of \mbox{633 m s$^{-1}$} was removed from the resulting spectrum.  The original ATMOS spectrum was apodized using a medium Beer-Norton function \citep{ATMOS94, Gunson}.  We hence applied the same apodization to our model output; this essential procedure is described in detail in Appendix~\ref{ATMOSappendix}.

To compute CO line bisectors, the absorption profiles were split at their centres, with each side separately interpolated to a 0.01 normalised intensity unit resolution scale using cubic splines.  Bisectors were then calculated as the series of wavelength midpoints between the two profile halves.  Wavelength errors were estimated conservatively for each bisector point from the previously published (wavelength-dependent) signal-to-noise of the data \citep{ATMOS96}, rather than the higher values we measured from the atlas .  Bisectors were truncated above 0.98 normalised intensity, as closer to the continuum they are often dominated by observational noise rather than line asymmetry.

\begin{figure*}[p]
\centering
\begin{narrow}{0in}{0in}
\begin{minipage}[c]{.46\linewidth}
	\centering
	\includegraphics[width=.88\textwidth]{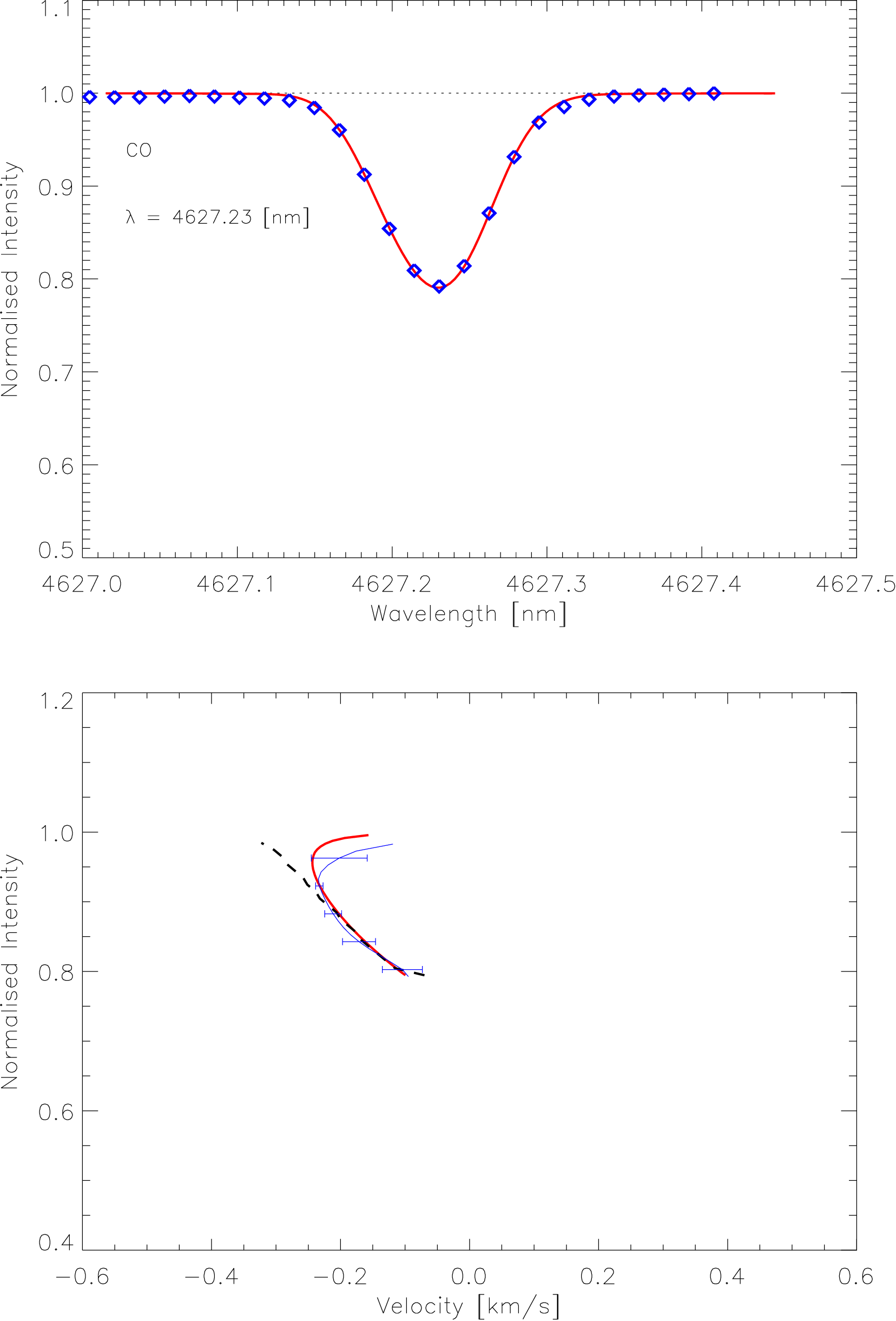}
	\\[5mm]
	\includegraphics[width=.88\textwidth]{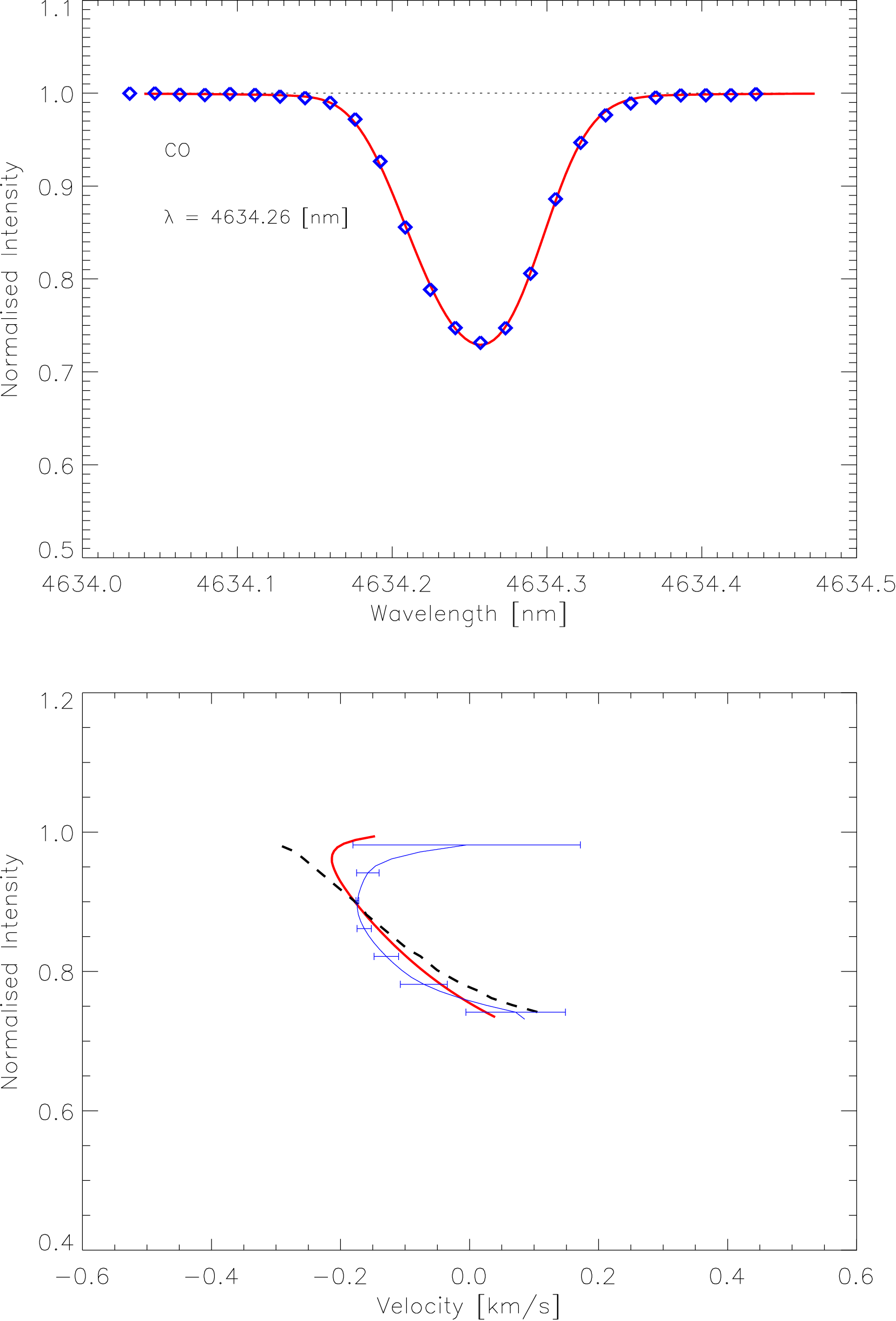}
\end{minipage}%
\hspace{0.08\linewidth}
\begin{minipage}[c]{.46\linewidth}
	\centering
	\includegraphics[width=.88\textwidth]{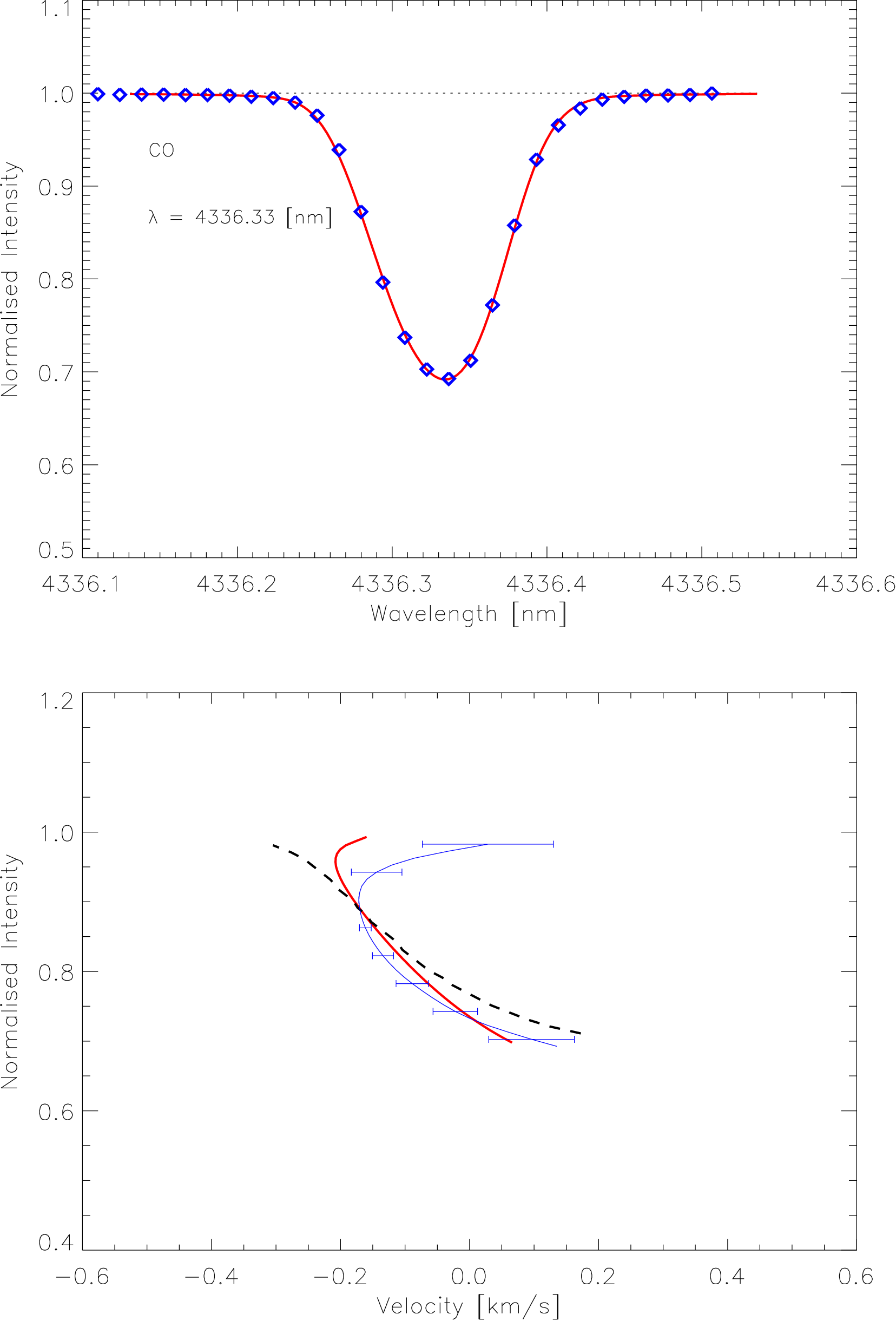}
	\\[5mm]
	\includegraphics[width=.88\textwidth]{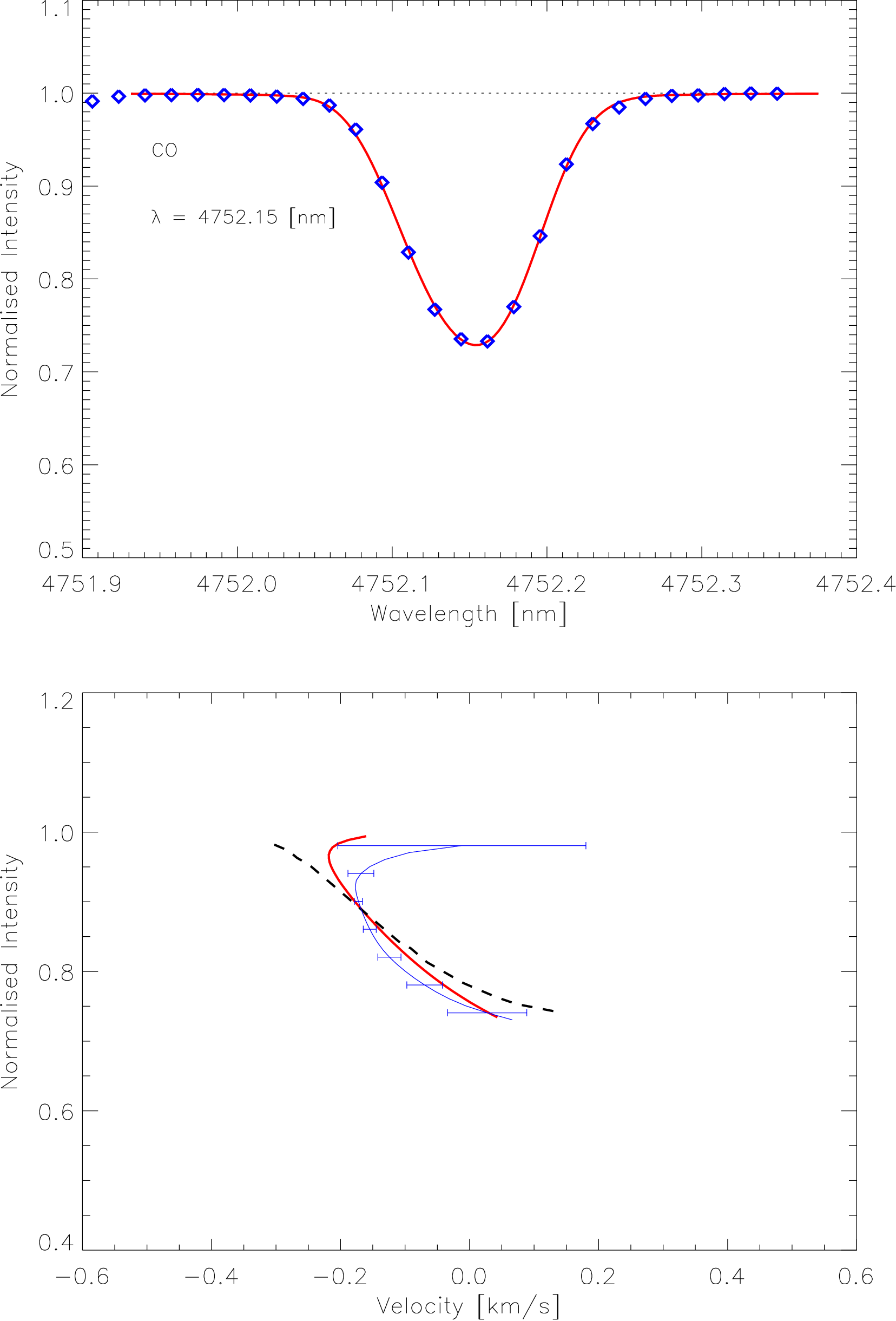}
\end{minipage}%
\caption[Example CO line profiles and bisectors]{Example spatially and temporally averaged, disk-centre synthesised strong CO line profiles and bisectors (solid lines without error bars), shown in comparison to ATMOS profiles (diamonds) and bisectors (solid lines with error bars).  The profile agreement is quite good, although as discussed in the text this requires an unreasonably high C and/or O abundance; the bisectors based on the new 3D model (see text) show better agreement with observation than those of the original 3D model version (dashed lines).  The solar gravitational redshift was removed from the ATMOS spectrum, and synthesised profiles have been apodized and fitted in abundance and Doppler shift.}
\label{bisectors}
\end{narrow}
\end{figure*}

\section{CO Line Shapes and Asymmetries}
\label{coresults}

\subsection{Analysis and Results}
\label{bsanalysis}

In this section, we restrict the discussion to the strongest CO lines, formed in the very high atmospheric layers embroiled in the current temperature debate \citep[e.g.][and references therein]{Ayr2002}.  Given their extreme formation heights, such lines are not ideal abundance determinants, as our study below demonstrates. In Sect. \ref{abunresults}, we instead present C and O isotopic abundances based on weaker CO lines formed in slightly lower layers, which our 3D hydrodynamic simulation reproduces more faithfully.

Modelled profiles were fitted in wavelength shift and abundance by minimising the $\chi^2$ likelihood estimator.  The wavelength fitting was permitted to allow for the possibility of systematic errors in the ATMOS wavelength calibration and/or the laboratory wavelengths of the lines considered.  Unfortunately, unfitted modelled line strengths were consistently weaker than those observed, requiring that we fit an unrealistically high carbon abundance to produce the best profile agreement.  Whilst resultant strong CO line profiles showed reasonable agreement with ATMOS profiles, their bisectors were of a $\backslash$-shape (Fig.~\ref{bisectors}, dashed lines) rather than the $\subset$-shape observed (solid lines with error bars).

The hydrodynamic simulations have undergone some slight improvements since the original model atmosphere was generated in 1999, so in an attempt to improve agreement we regenerated the CO line profiles using the most modern version of the simulations available.  The newer code contained an improved treatment of the numerical viscosity, and drew on an updated Uppsala package containing slightly improved continuous opacities.  The version of the MHD equation-of-state was also more recent, including argon in addition to the 16 other most abundant elements, as well as drawing on slightly altered abundances.  The manner in which radiation pressure and energy were included by the MHD tabulation program was also slightly altered.  Finally, the mean density and temperature structures used in each case to generate equation-of-state and opacity tables were slightly different, though it is not clear whether this was a primary difference or a secondary effect caused by other updates.  The final temperature structures were very nearly identical, however.

Inspection suggested that the failure to properly reproduce strong lines might have been due to incoming gas from the uppermost layer not having time to properly cool and equilibrate to its surroundings before it reached CO line formation heights.  To test this hypothesis, a version of the new atmosphere extending to heights of almost \mbox{1.2 Mm} was also created.  The extended simulations were run at a resolution of \mbox{50 $\times$ 50 $\times$ 88} and those with the original extension at \mbox{50 $\times$ 50 $\times$ 82}.  The lower resolutions were chosen due to computational time constraints, as the purpose was not to produce the most accurate description of the solar atmosphere available, but to provide a qualitative idea of any improvement in CO bisectors due to the later code or boundary extension.  This is an appropriate approach for the qualitative analyses carried out in this Section, but not for e.g. abundance analyses, as carried out in Sect.~\ref{abunresults}.

Profiles were simulated for all 31 lines, apodized and fitted.  Using the regenerated 3D model atmosphere, very good agreement was obtained with observed profiles and bisectors.  Unfitted line strengths were still smaller than observed, but the discrepancy was much less than seen previously.  Fitted abundances were hence still unrealistically high, though the discrepancy with abundances from weak lines (derived in Sect.~\ref{abunresults}) was only about half that produced by the original model atmosphere.  The scatter in abundances from strong lines was also halved when the regenerated model was employed.  Some example resultant profiles and bisectors from the (non-extended) regenerated model atmosphere are shown in comparison to their ATMOS counterparts in Fig.~\ref{bisectors}.  Bisectors derived with the older version of the model atmosphere are also given for comparison.  The average reduced $\chi^2$ value for the agreement between observed and theoretical profiles was an order of magnitude smaller using the later model atmosphere.  Profiles and bisectors derived with the extended simulation showed no differences to those produced with the corresponding non-extended version (data not shown).  We hence focus mainly on the results from the unextended, later version of the model atmosphere for the remainder of our analysis of line shapes and asymmetries.

\subsection{Discussion}
\label{bsdiscussion}

To our knowledge, the only prior study of observed asymmetries in solar CO lines is that of \citet{GSbi}.  Without even considering modelled profiles and bisectors, it is plain that the observed ATMOS bisectors do not agree with the $\backslash$-shaped bisectors found previously, a significant result in itself.  The reason for this discrepancy appears to simply be that \citet{GSbi} extended their bisectors only to intensities of about 0.94 of the continuum intensity, whereas those of Fig.~\ref{bisectors} extend to intensities of 0.98.  Given that the bisectors of \citet{GSbi} were taken from the more noisy SL-3 flight of the ATMOS instrument rather than the ATLAS-3 flight, this lesser extension is not surprising.

The hydrodynamic simulations appear to have been sufficiently improved for some of the deficiencies in the original to be removed.  Speculating exactly which minute change or combination of changes described in Sect.~\ref{bsanalysis} is responsible for such a higher-order effect as the turning of one end of a bisector is exceedingly difficult.  All that can confidently be said on this matter is that it is very unlikely the introduction of argon to the equation-of-state had any effect.  Beyond this, any combination of the tiny changes in opacity and abundance inputs, viscosity and radiation pressure treatments or initial mean structure could be responsible.  

However, whether the improvement is attributable to resolutional effects rather than actual model differences must be considered.  \citet{AspRes} showed that there is an increase in strong Fe line width going from the resolution of the later model to that of the original, with a preferential filling in of the bluewards wing of profiles over the redwards wing.  This is due to the asymmetric way in which low resolution models truncate the vertical velocity distribution, removing more of the high upwards than downwards velocities.  This correspondingly manifests itself in a bluewards turning of bisector tops with higher resolution, or equivalently, as a redwards `turning-back' of bisectors with insufficient resolution.

This `turning-back' has certainly occurred in the current study, so the important question is not whether, but to what extent the improved bisectors and profiles are attributable to the lower resolution.  The bisectors of \citet{AspRes} show a maximum redshifting of the bisector top of \mbox{150 m\,s$^{-1}$} with the change from a \mbox{200 $\times$ 200 $\times$ 82} to \mbox{50 $\times$ 50 $\times$ 82} model.  They also show a redwards shifting of bisector feet of approximately \mbox{100 m\,s$^{-1}$}, resulting in a net redwards movement of \mbox{50 m\,s$^{-1}$} of the bisector top relative to the foot.  In contrast, the newer model bisectors demonstrate a bare minimum of \mbox{250 m\,s$^{-1}$} net shift of the top relative to the foot, compared with the (dashed) originals.  It would also seem a rather unlikely coincidence for the tenfold difference in $\chi^2$ values to be due purely to resolutional effects.  The improvement of profile shapes with the regenerated model over the original version is also borne out by a much improved convergence in fitted line depths and strengths.  Fitted abundances from the 31 lines show far less scatter in the case of the later model, and actual abundance values are also closer to those derived later using weak lines.  These changes support the notion that resolutional effects do not play a dominant role in the bisector and profile improvements, as the effects of reduced resolution are expected to be the \emph{opposite} of that observed, i.e. an increase in scatter \citep{AspRes} and abundances derived from CO lines (this paper, Sect.~\ref{cdiscuss}).  On the basis of these differences, we estimate the contribution of resolutional effects to be approximately half of the improvement seen.

Whilst agreement has been improved, the final results indicate that the description of the uppermost layers in the model atmosphere is still not perfect. This is not surprising, as many of the physical approximations used begin to break down in the highest layers.  Processes such as non-equilibrium hydrogen ionisation, non-LTE continuum formation and extended CO cooling related to non-equilibrium chemistry might actually be required for an accurate description of such regions.  Ignoring these non-equilibrium processes could lead to slightly anomalous velocity or temperature structures in the top layers, issues discussed in more depth in Sect.~\ref{cdiscuss}.  Indeed, it turns out that the most likely problem is a slight overestimation of temperature in the uppermost layers of the model, possibly more so in intergranular lanes than granules.  This is indicated by the overly high abundances derived from strong and LE lines (as will be seen in Sect.~\ref{cdiscuss}), as well as our own more extensive internal investigations into the bisector discrepancies and the regeneration exercise (data not shown).  Though imperfect, the agreement that was seen with observed bisectors does however indicate that the 3D model at least captures some of the essential physical features of these very high layers, which is encouraging.

\subsection{Temperature Structure and CO Distribution}
\label{tempdiscuss}

To use a model to draw accurate conclusions about the temperature structure of the solar atmosphere and the distribution of CO within it, one would ideally require that synthesised CO lines demonstrate excellent agreement with observations from very high atmospheric layers.  Unfortunately, despite being the best effort to date, this has not been the case in this study.  However, considering that we were able to at least partially reproduce such difficult indicators as the strong line bisectors, and more importantly, that all our investigations point towards the mean temperature structure in the upper layers of the model atmosphere being slightly too high, we can use the existing 3D model atmosphere to indicate temperature upper bounds for the disputed layers.  The temperature structure of Fig.~\ref{formationheights} hence suggests that cool gas does indeed exist in the lower chromosphere, with an average temperature of $<$4000K.  Such an upper bound is consistent with the extended 3D LTE atmosphere of \citet{wedemeyer04}, and the 3500K proposed by \citet{Ayr2002}.  Furthermore, the temperature at the site of the previous `minimum' also seems lower than previously thought, at $<$4200K rather than $\sim$4500K.  Owing to the inhomogeneity of the model utilised here, we also tentatively suggest the existence of some persistent gas with $T<3700$K at COmospheric heights, and even intermittent gas temperatures of less than 2000K.  

The lack of any profile or bisector improvement in the extended simulations would seem to indicate that the extra layers above \mbox{1 Mm} probably do not contribute to CO line formation at disc centre for the lines considered here.  This also suggests that the downflowing gas entering the upper layer of the non-extended simulations does not play a role in the line formation until it has passed significantly beyond the top boundary layer and been permitted to adjust to the temperature of its surroundings.  Hence, gas above a height of \mbox{$\sim$0.75 Mm} might be identified as also generally not contributing to disk-centre solar CO line formation (except perhaps for some extremely strong lines).  Furthermore, as it will take some distance to achieve temperature equilibration once gas has passed out of the topmost layer of the simulation, our results could be seen to suggest an uppermost effective extent of the COmosphere of around \mbox{700 km}, consistent with the indications of \citet{AR2003}.  It should be noted that this would not however preclude the existence of cool(er) gas at greater heights, just suggests an uppermost extent of CO in \emph{significant} concentrations.

\section{C and O Isotopic Abundance Determinations}
\label{abunresults}

\subsection{Dealing with CO and Isotopes}
\label{dealing}

Working with a known oxygen abundance, a fitted abundance difference for CO lines can be approximately interpreted as a difference in solar carbon abundance, even though strictly speaking it is actually indicative of a difference in CO abundance at the height of line formation.  In order to use CO lines to derive a solar carbon abundance, the input carbon abundance was therefore iteratively altered according to the average fitted abundances of the CO lines in the previous iteration, until convergence was obtained.  Oxygen concentration effects the equilibrium position of the CO formation-dissociation reaction, so carbon abundances derived in this manner are dependent upon the adopted oxygen abundance.  Input oxygen abundances were hence carefully chosen (cf.~Sect.~\ref{model}) and kept constant throughout the study.

Using the above technique, the solar $^{12}$C abundance was determined separately using the three different $^{12}$C$^{16}$O line lists.  Since the 3D model input abundances do not differentiate between isotopes, these determinations were performed by altering the total input carbon abundance, though the final abundance arrived at is not overall abundance, but the carbon abundance \emph{if all C in CO were contained in $^{12}$C$^{16}$O}.  This can approximately be called the $^{12}$C abundance, even though it does not include any $^{12}$C tied up in $^{12}$C$^{17}$O or $^{12}$C$^{18}$O.  Though they are minimal given the high values of $^{16}$O/$^{18}$O and $^{16}$O/$^{17}$O, these contributions were taken into account in the calculation of a total carbon abundance by the use of a fractional scalefactor (Appendix~\ref{fracsf}).

The concentrations of $^{13}$C$^{16}$O and $^{12}$C$^{18}$O relative to $^{12}$C$^{16}$O are identical to the $^{12}$C/$^{13}$C and $^{16}$O/$^{18}$O ratios.  In order to determine the concentrations of the CO isotopomers, given the lack of provision for isotopic differentiation in the model, the isotopomeric lines were treated in the radiative transfer simulations as if they were created by $^{12}$C$^{16}$O.  In order to do this correctly, the radiative transfer code had to be altered to include provision for a mass scalefactor and an opacity scalefactor (refer to Appendix~\ref{scalefactors}).

\subsection{Abundance Calculations}
\label{abuncalcs}

For each iteration, three profiles with $\log gf$ values differing by \mbox{0.2 dex} were calculated for a given line and interpolated between using cubic splines to arrive at the desired line strength.  As described in Sect.~\ref{model}, abundance calculations were carried out for all five line lists with four different models: 3D, HM, \textsc{marcs} and 1DAV.  The 3D model used was that of \citet{AspI}, the same model atmosphere used for all previous 3D abundance determinations \citep[e.g.][]{AspII, AspIV, AspRev, AspVI, AspIII, AspV}.  Since accuracy is paramount in abundance determinations, the regenerated version described in Sect.~\ref{coresults} would have been inappropriate given its reduced resolution.  This model was also used over the regenerated version for consistency and comparability with the previous abundance determinations.

Both equivalent width and profile fitting (the latter via $\chi^2$-analysis) were tested upon the first iteration of the weak $^{12}$C$^{16}$O lines, with virtually no difference found in calculated abundance.  However, some of the lines could not be effectively profile fitted by minimising the $\chi^2$ statistic without some form of masking, due to the presence of other nearby lines in the ATMOS spectrum.  For this reason, equivalent width fitting was used for all subsequent abundance measures, as this problem was only likely to worsen when the weaker isotopomeric lines were considered.  The same Beer-Norton apodization as used in Sect.~\ref{coresults} (medium BNA, characteristic velocity \mbox{1.5 km\,s$^{-1}$}) was also applied to each line, though more for the sake of consistency than anything else.  Convolution with any function normalised to unit area (as the medium Beer-Norton used was) should be an area-preserving operation, thereby not effecting line equivalent widths nor therefore the abundances determined with them.  For completeness however, it should be noted that in the case of the \mbox{$\Delta v = 2$} lines, a BNA characteristic velocity of \mbox{0.75 km\,s$^{-1}$} was used, reflecting the higher resolving power of the ATMOS instrument at the shorter wavelengths of these lines (see Table~\ref{dv2lines}).

\begin{table*}[tbp]
\centering
\caption[Abundances and isotopic ratios implied by all $^{12}$C$^{16}$O lines]{Abundances and isotopic ratios implied by the weak (\mbox{$\Delta v = 1$}), LE (\mbox{$\Delta v = 1$}) and \mbox{$\Delta v = 2$} $^{12}$C$^{16}$O line lists as indicator of $^{12}$C abundance.  Note the large differences between 1D and 3D isotopic ratios.  Note also that the HM and \textsc{marcs} 1D models indicate higher abundances than the 1DAV, which in turn produces a higher abundance than the 3D model.}
\label{tableCO}
\begin{tabular}{l l c c c c c}
\hline
$^{12}$C$^{16}$O Lines & Model & 3D & HM & \textsc{marcs} & 1DAV & Terrestrial\\
\hline
Weak & $\log \epsilon_\mathrm{C}$ & $8.40\pm0.01$ & $8.60\pm0.01$ & $8.55\pm0.02$ & $8.46\pm0.01$ & \\
($\Delta v = 1$) & $\log \epsilon_{^{12}\mathrm{C}}$ & $8.39\pm0.01$ & $8.59\pm0.01$ & $8.55\pm0.02$ & $8.45\pm0.01$ & \\
& $\log \epsilon_{^{13}\mathrm{C}}$ & $6.44 \pm 0.02$ & $6.76 \pm 0.02$ & $6.70 \pm 0.02$ & $6.58 \pm 0.02$ & \\
& $\log \epsilon_{^{18}\mathrm{O}}$ & $5.97 \pm 0.03$ & $6.31 \pm 0.04$ & $6.25 \pm 0.04$ & $6.12 \pm 0.03$ & \vspace{0.5mm}\\
& $^{12}$C/$^{13}$C & 88.8$^{+5.3}_{-5.0}$ & 68.6$^{+3.2}_{-3.0}$ & 70.6$^{+4.3}_{-4.1}$ & 74.0$^{+3.3}_{-3.2}$ & $89.4\pm 0.2$ \vspace{1mm}\\
& $^{16}$O/$^{18}$O & 490$^{+41}_{-37}$ & 344$^{+30}_{-28}$ & 350$^{+35}_{-32}$ & 376$^{+31}_{-29}$ & $498.7\pm 0.1$ \vspace{1mm}\\
\hline
LE & $\log \epsilon_\mathrm{C}$ & $8.48\pm0.04$ & $8.66\pm0.02$ & $8.67\pm0.03$ & $8.49\pm0.02$ & \\
($\Delta v = 1$) & $\log \epsilon_{^{12}\mathrm{C}}$ & $8.47\pm0.04$ & $8.66\pm0.02$ & $8.67\pm0.03$ & $8.49\pm0.02$ & \\
& $\log \epsilon_{^{13}\mathrm{C}}$ & $6.44 \pm 0.02$ & $6.76 \pm 0.02$ & $6.70 \pm 0.02$ & $6.58 \pm 0.02$ & \\
& $\log \epsilon_{^{18}\mathrm{O}}$ & $5.89 \pm 0.03$ & $6.24 \pm 0.04$ & $6.13 \pm 0.04$ & $6.09 \pm 0.03$ & \vspace{0.5mm}\\
& $^{12}$C/$^{13}$C & 107.6$^{+13}_{-12}$ & 80.3$^{+5.8}_{-5.4}$ & 93.2$^{+8.7}_{-7.9}$ & 80.0$^{+5.1}_{-4.8}$ & $89.4\pm 0.2$ \vspace{1mm}\\
& $^{16}$O/$^{18}$O & 594$^{+80}_{-71}$ & 403$^{+42}_{-38}$ & 463$^{+57}_{-50}$ & 407$^{+39}_{-35}$ & $498.7\pm 0.1$ \vspace{1mm}\\
\hline
\mbox{$\Delta v = 2$} & $\log \epsilon_\mathrm{C}$ & $8.37\pm0.01$ & $8.69\pm0.02$ & $8.58\pm0.02$ & $8.51\pm0.02$ & \\
& $\log \epsilon_{^{12}\mathrm{C}}$ & $8.36\pm0.01$ & $8.68\pm0.02$ & $8.57\pm0.02$ & $8.50\pm0.02$ & \\
& $\log \epsilon_{^{13}\mathrm{C}}$ & $6.44 \pm 0.02$ & $6.76 \pm 0.02$ & $6.70 \pm 0.02$ & $6.58 \pm 0.02$ & \\
& $\log \epsilon_{^{18}\mathrm{O}}$ & $6.00 \pm 0.03$ & $6.22 \pm 0.04$ & $6.23 \pm 0.04$ & $6.07 \pm 0.03$ & \vspace{0.5mm}\\
& $^{12}$C/$^{13}$C & 82.8$^{+5.2}_{-4.9}$ & 83.9$^{+5.9}_{-5.5}$ & 75.0$^{+5.7}_{-5.3}$ & 82.8$^{+4.5}_{-4.3}$ & $89.4\pm 0.2$ \vspace{1mm}\\
& $^{16}$O/$^{18}$O & 457$^{+39}_{-36}$ & 421$^{+43}_{-39}$ & 372$^{+41}_{-37}$ & 421$^{+38}_{-35}$ & $498.7\pm 0.1$ \vspace{1mm}\\
\hline
\end{tabular}
\end{table*}

\begin{figure*}[tbp]
\centering
\begin{narrow}{0in}{0in}
\begin{minipage}[c]{.44\linewidth}
	\includegraphics[width=\textwidth]{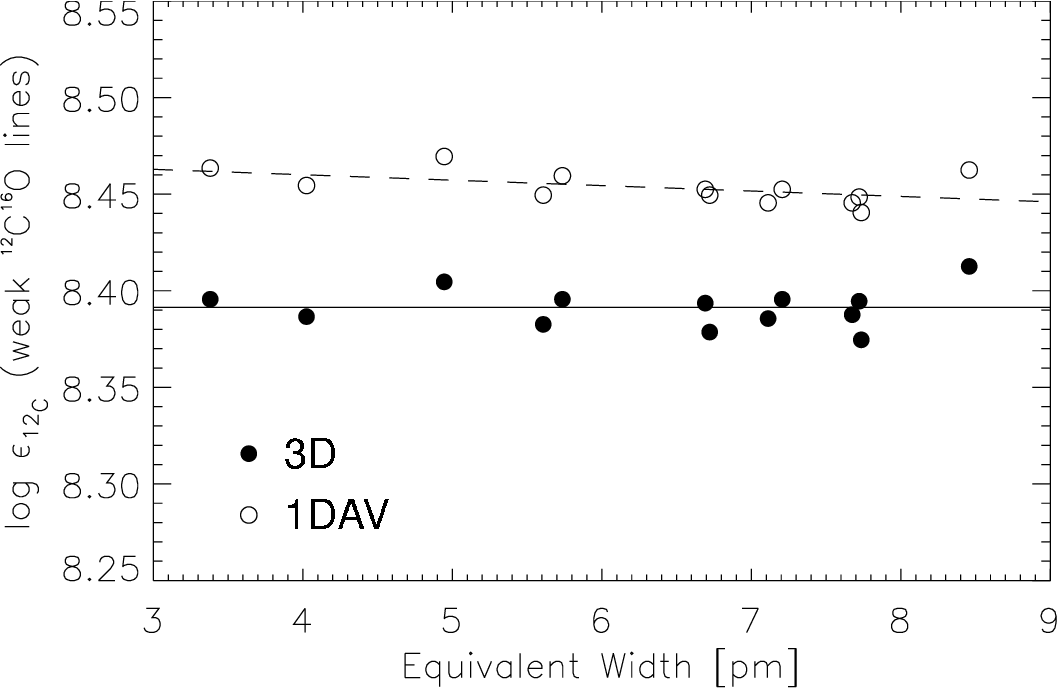}
	\\[5mm]
	\includegraphics[width=\textwidth]{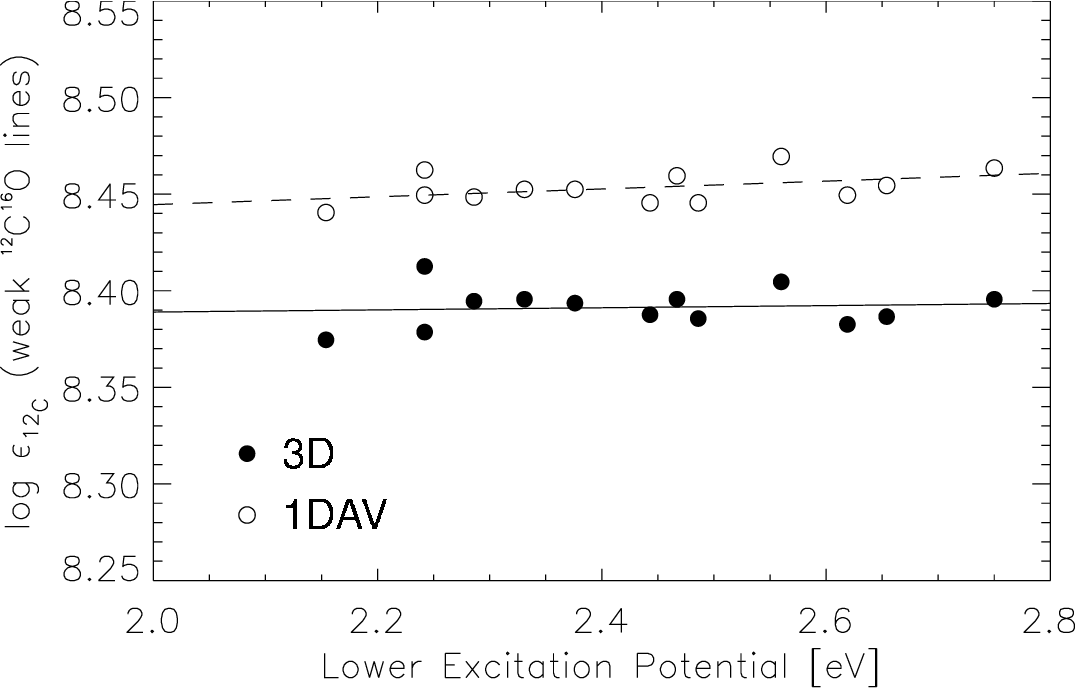}
\end{minipage}%
\hspace{0.12\linewidth}
\begin{minipage}[c]{.44\linewidth}
	\includegraphics[width=\textwidth]{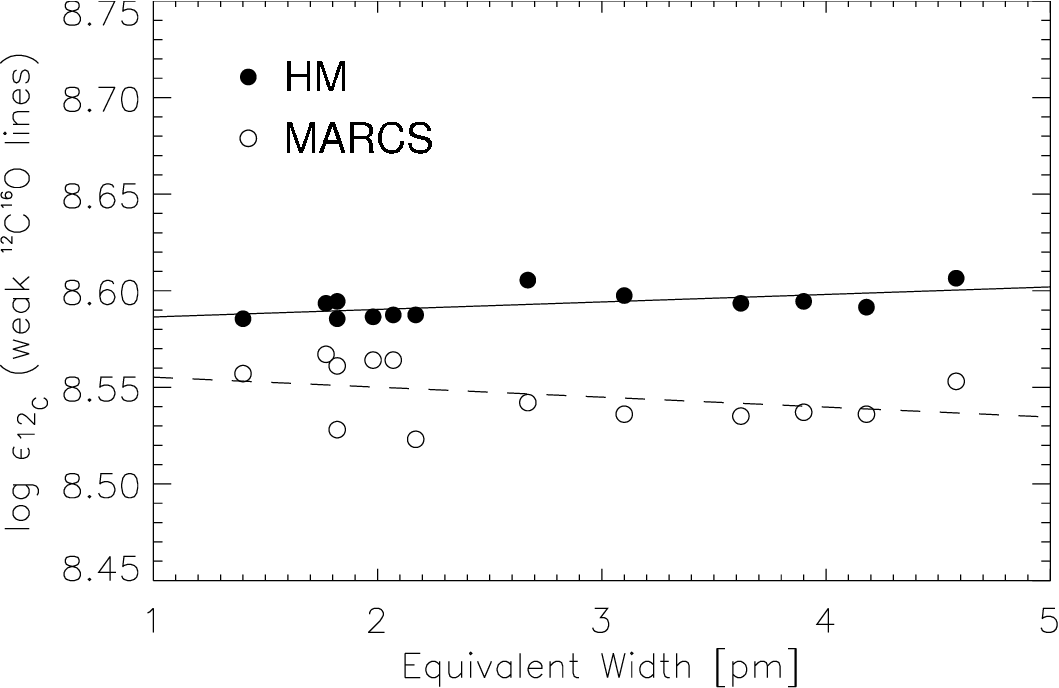}
	\\[5mm]
	\includegraphics[width=\textwidth]{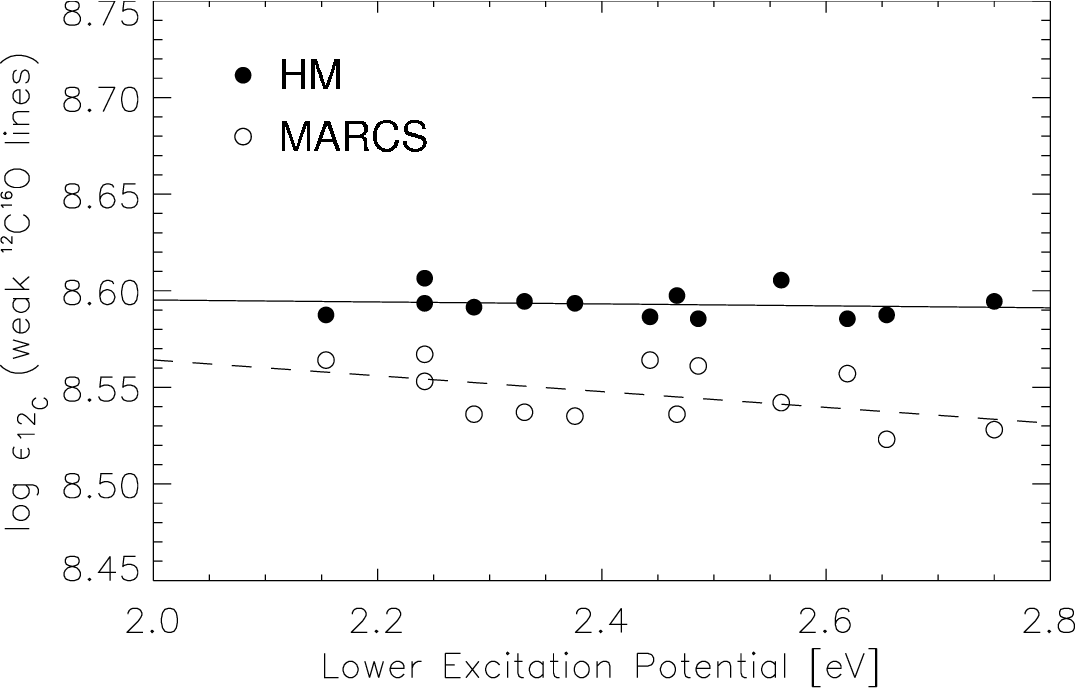}
\end{minipage}%
\caption[Weak $^{12}$C$^{16}$O-indicated $^{12}$C abundance plots]{Solar carbon abundances indicated by the weak $^{12}$C$^{16}$O lines, displayed according to equivalent width (top) and excitation potential (bottom).  On the left, filled circles indicate 3D results and open circles 1DAV results, whilst on the right filled circles are HM values and open circles \textsc{marcs} results.  Trendlines are produced as linear fits to data sets using a minimised $\chi^2$ method placing equal weight on each point, with solid lines corresponding to filled circles and dashed lines to open circles.  No significant trends can be seen in the output of any model.}
\label{plotsCOweak}
\centering
\end{narrow}
\end{figure*}

\begin{figure*}[tbp]
\centering
\begin{narrow}{0in}{0in}
\begin{minipage}[c]{.44\linewidth}
	\includegraphics[height=50mm]{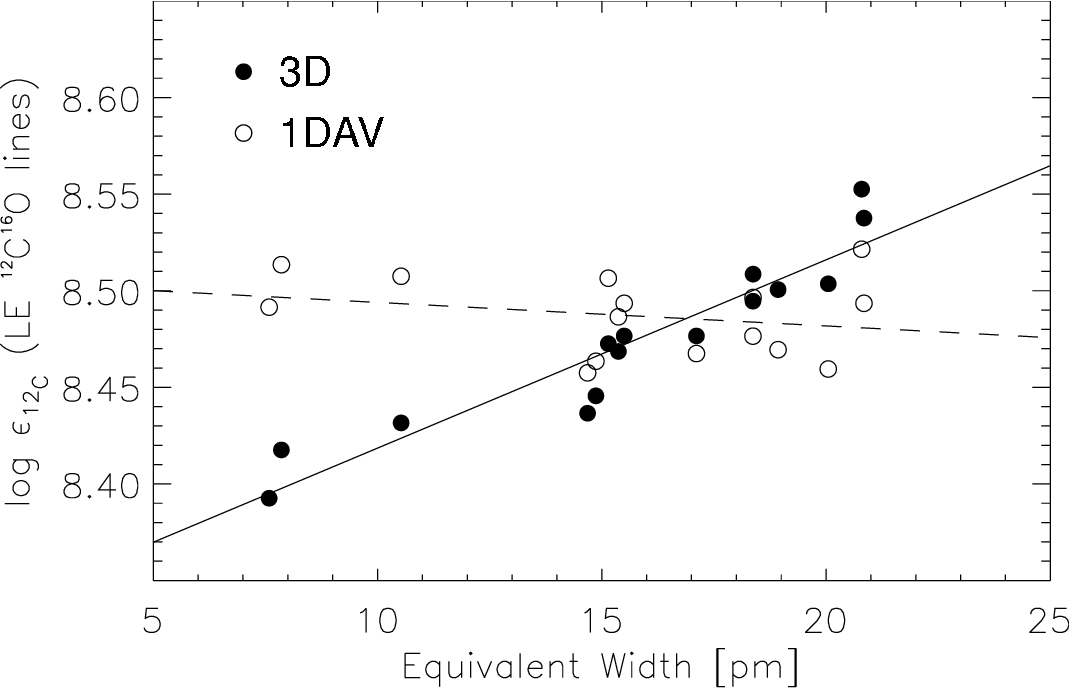}
	\\[5mm]
	\includegraphics[height=50mm]{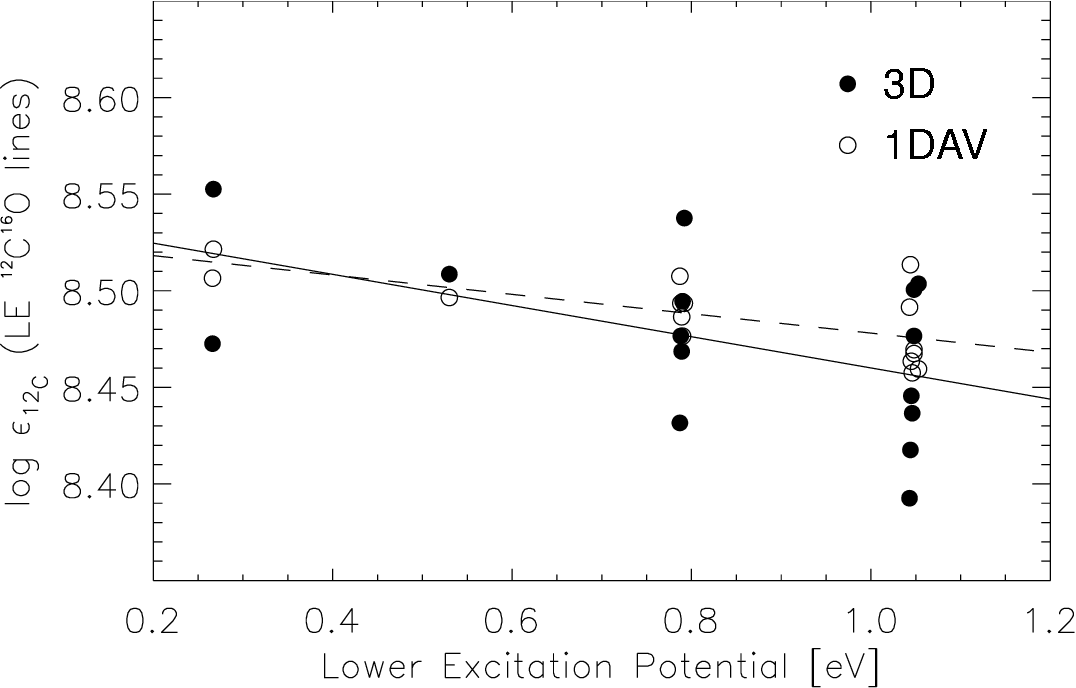}
\end{minipage}%
\hspace{0.04\linewidth}
\begin{minipage}[c]{.44\linewidth}
	\includegraphics[height=50mm]{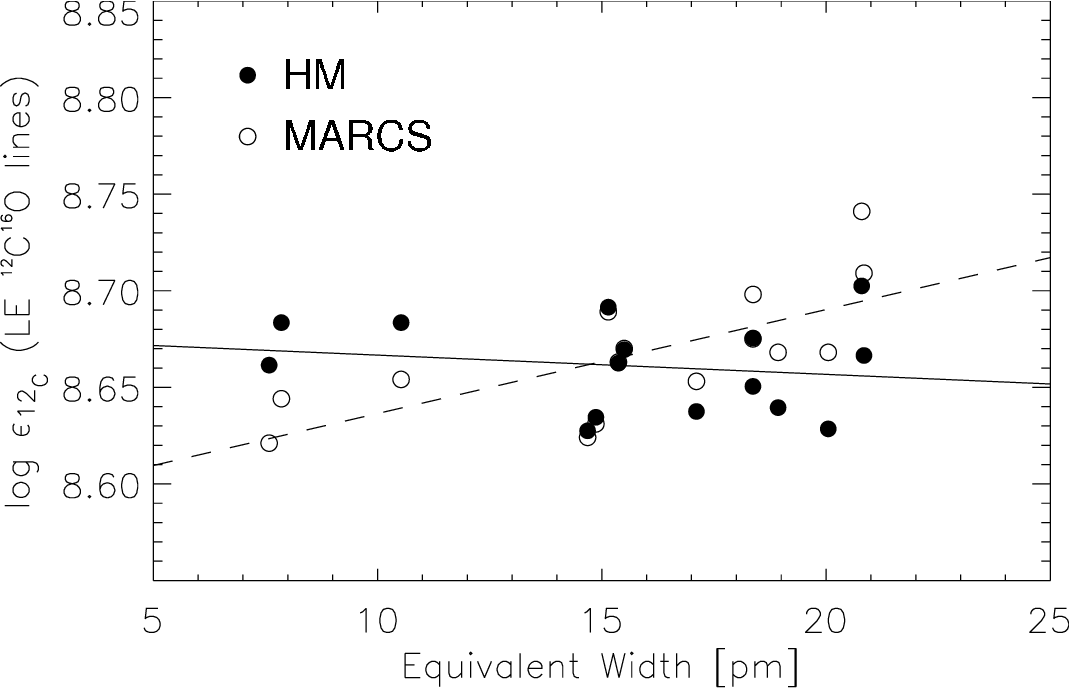}
	\\[5mm]
	\includegraphics[height=50mm]{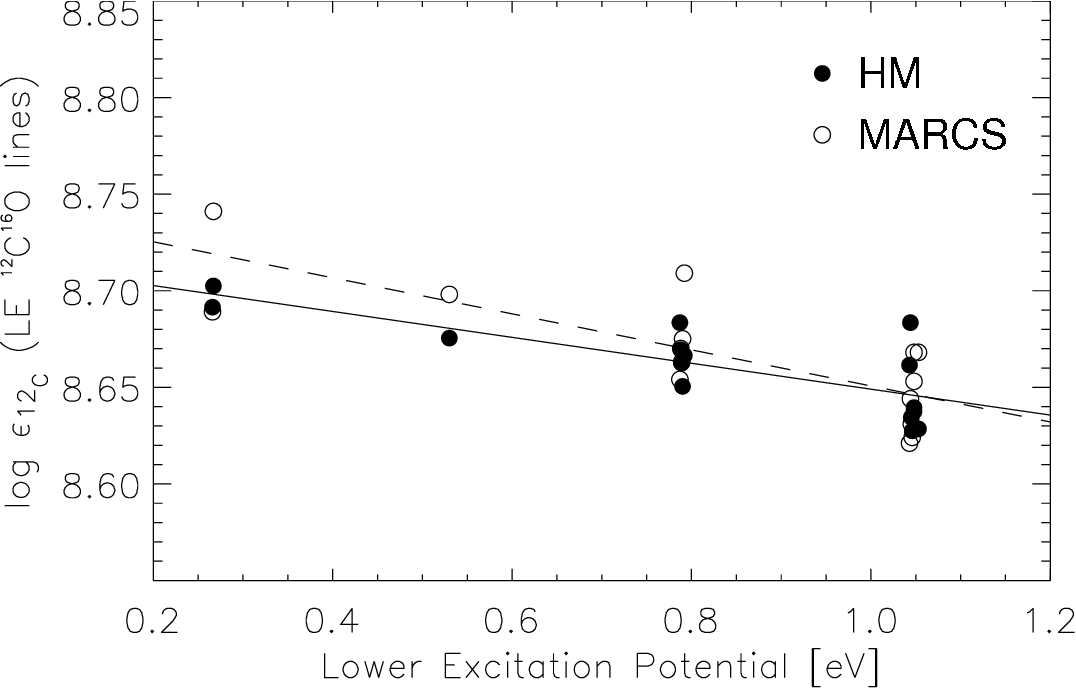}
\end{minipage}%
\caption[LE $^{12}$C$^{16}$O-indicated $^{12}$C abundance plots]{The same as Fig.~\protect\ref{plotsCOweak}, but using the LE $^{12}$C$^{16}$O lines.  Definite trends can be seen with equivalent width in the output of the 3D and \textsc{marcs} models.  Significantly, the $^{12}$C abundances implied by the weakest (i.e. lowest formation height) lines in 3D are consistent with the abundances derived using the weak and \mbox{$\Delta v = 2$} $^{12}$C$^{16}$O lines.  The agreement deteriorates with increased LE $^{12}$C$^{16}$O line strength and therefore formation height. Less prominent trends are also evident in excitation potential for all models.}
\label{plotsCOLE}
\centering
\end{narrow}
\end{figure*}

\begin{figure*}[tbp]
\centering
\begin{narrow}{0in}{0in}
\begin{minipage}[c]{.44\linewidth}
	\includegraphics[height=50mm]{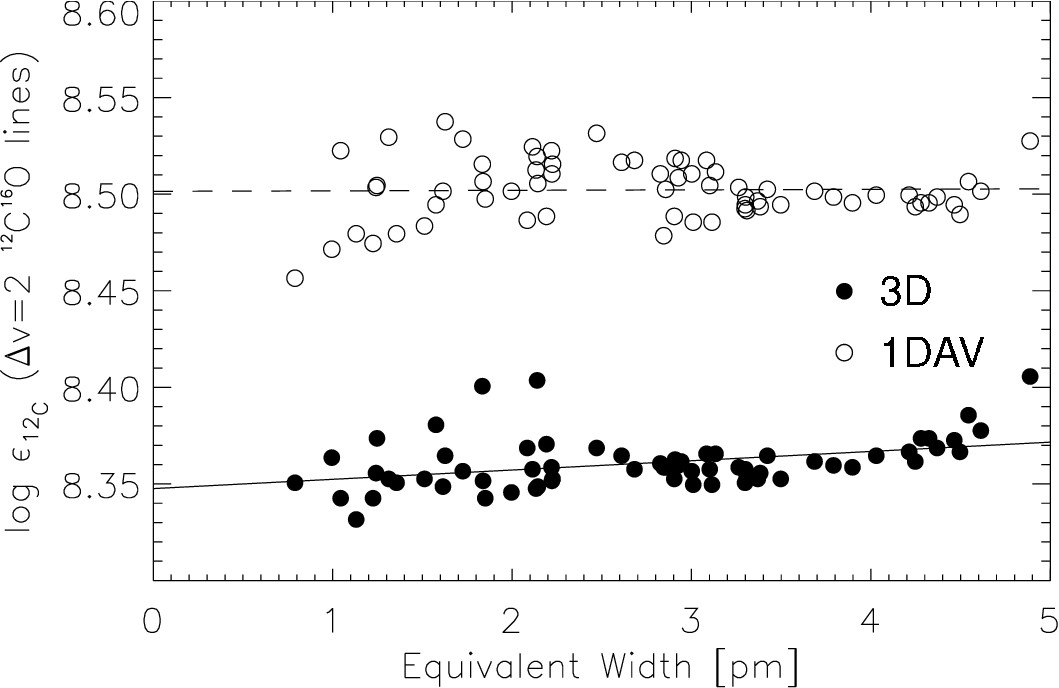}
	\\[5mm]
	\includegraphics[height=50mm]{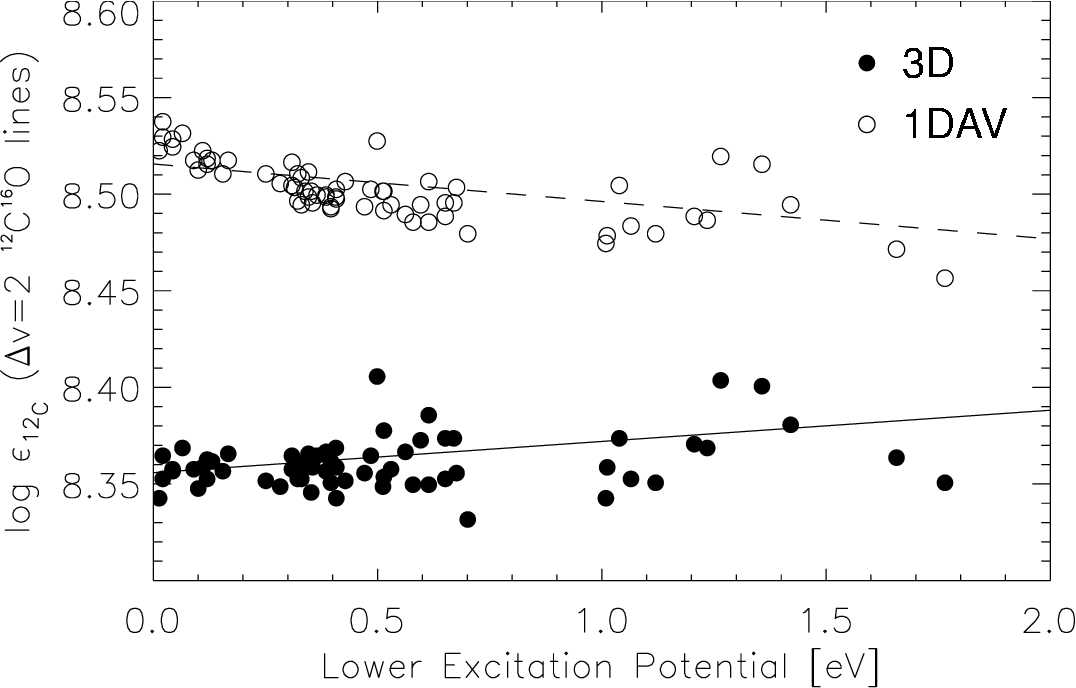}
\end{minipage}%
\hspace{0.12\linewidth}
\begin{minipage}[c]{.44\linewidth}
	\includegraphics[height=50mm]{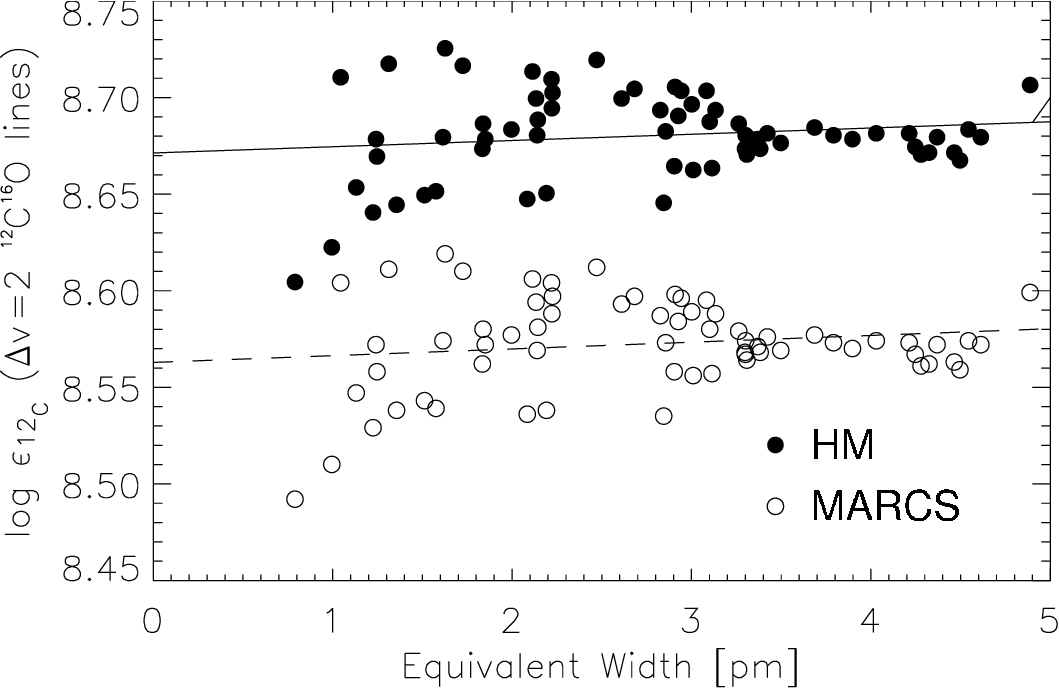}
	\\[5mm]
	\includegraphics[height=50mm]{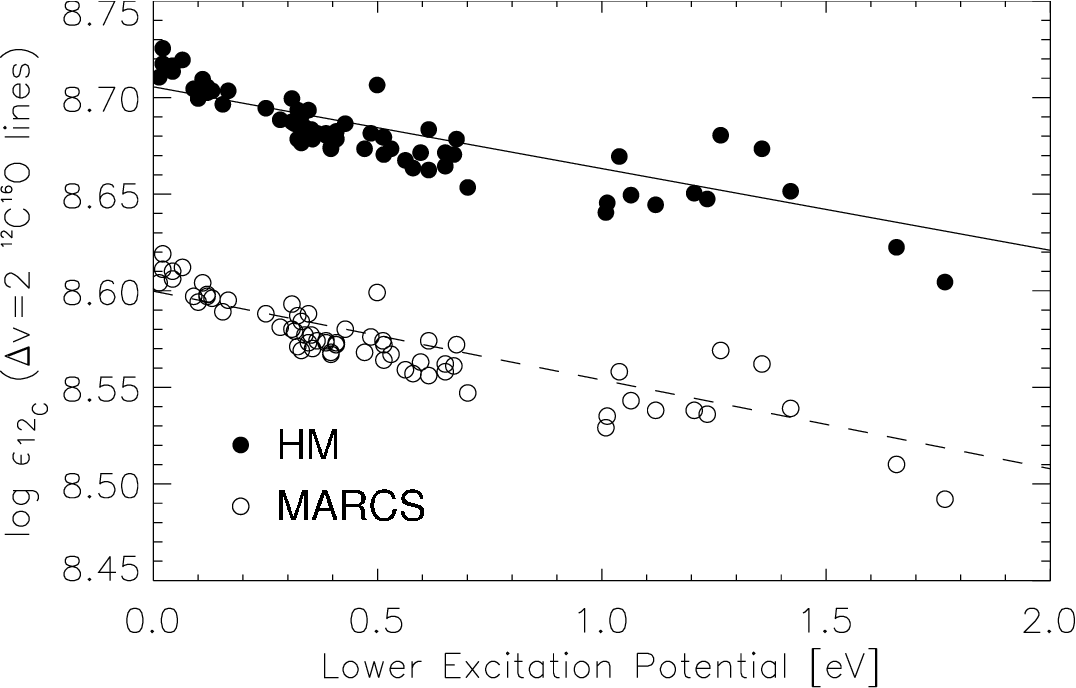}
\end{minipage}%
\caption[\mbox{$\Delta v = 2$} $^{12}$C$^{16}$O-indicated $^{12}$C abundance plots]{The same as Fig.~\protect\ref{plotsCOweak}, but using the \mbox{$\Delta v = 2$} $^{12}$C$^{16}$O lines.  Significant trends can be seen with excitation potential in the 1D results, though none in the case of the 3D model.}
\label{plotsCOdv2}
\end{narrow}
\end{figure*}

\begin{figure*}[tbp]
\centering
\begin{narrow}{0in}{0in}
\begin{minipage}[c]{.44\linewidth}
	\includegraphics[height=50mm]{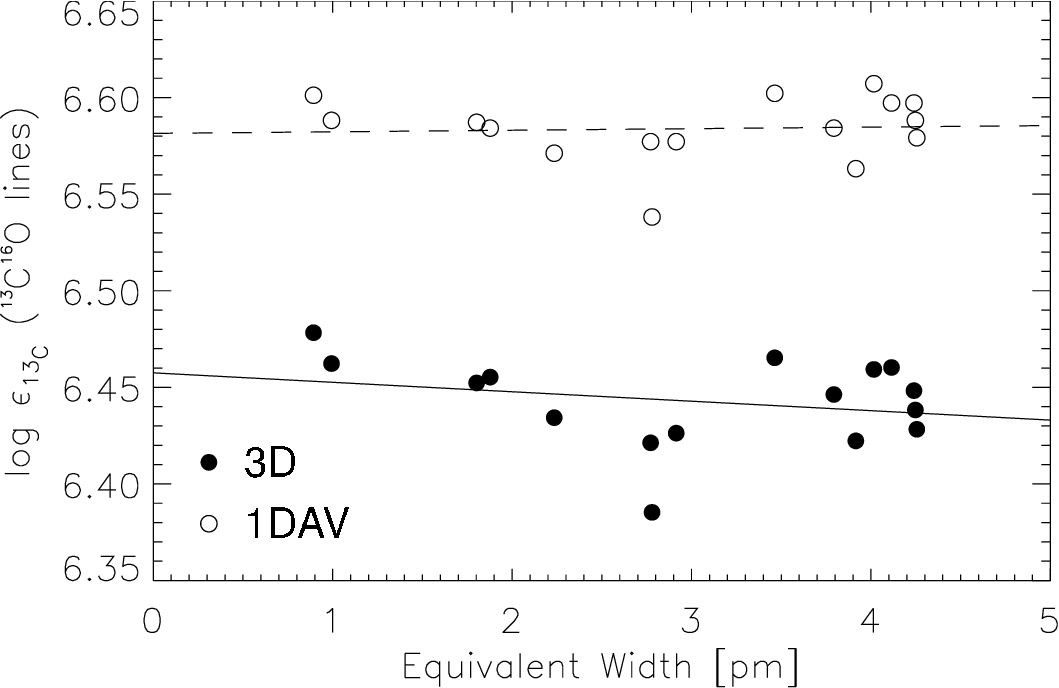}
	\\[5mm]
	\includegraphics[height=50mm]{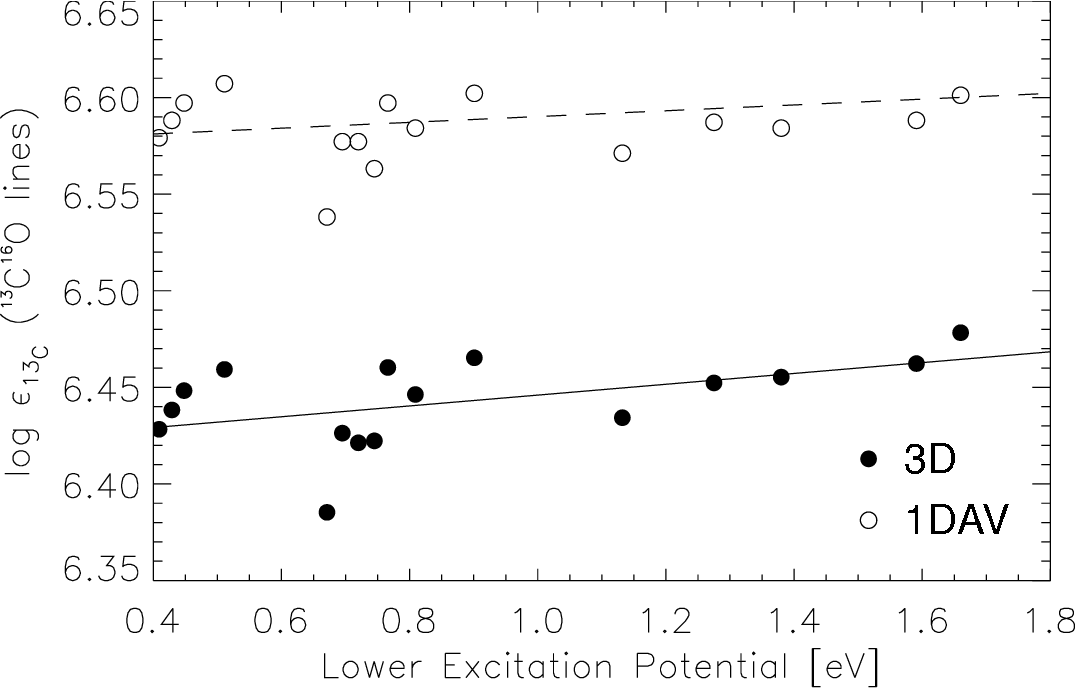}
\end{minipage}%
\hspace{0.12\linewidth}
\begin{minipage}[c]{.44\linewidth}
	\includegraphics[height=50mm]{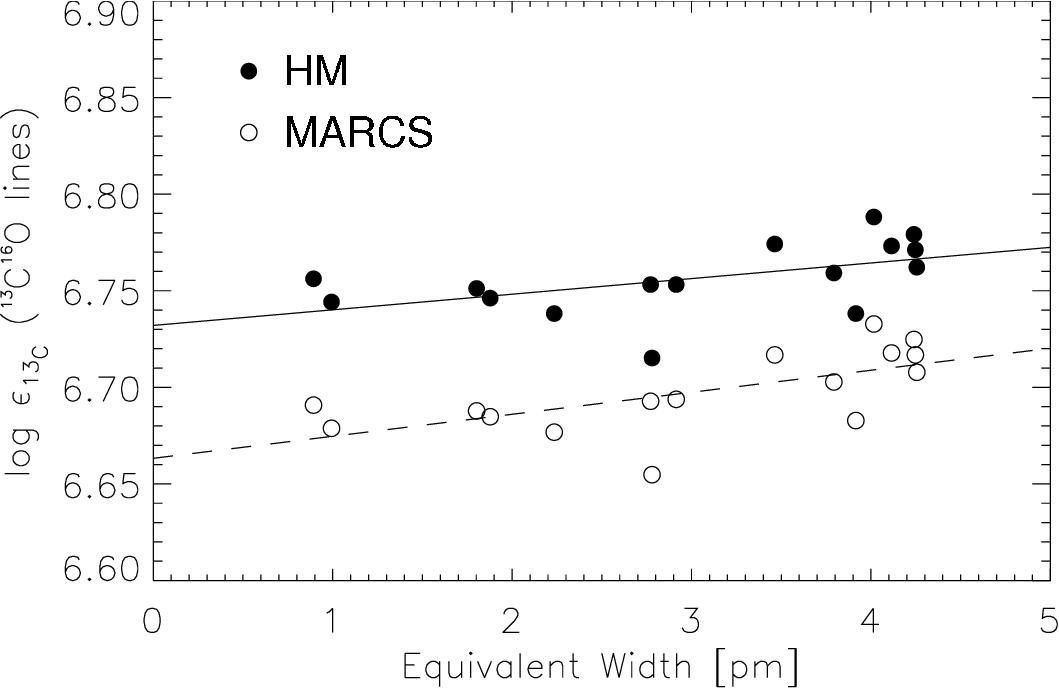}
	\\[5mm]
	\includegraphics[height=50mm]{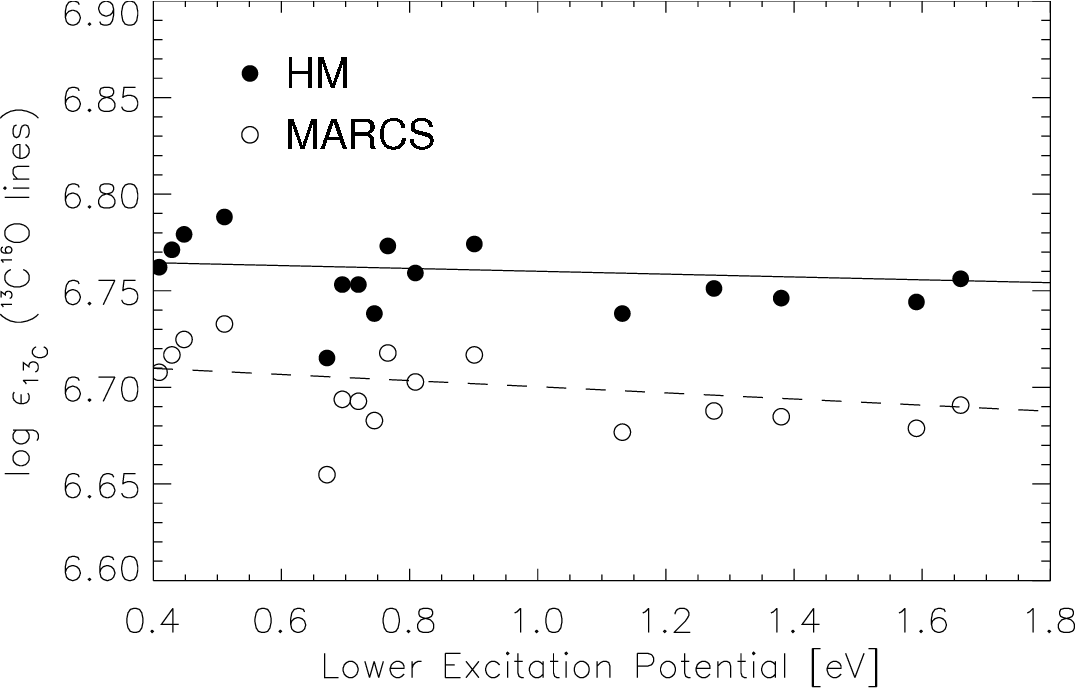}
\end{minipage}%
\caption[$^{13}$C abundance plots]{The same as Fig.~\protect\ref{plotsCOweak}, but indicating $^{13}$C through the use of the $^{13}$C$^{16}$O lines.  No significant trends can be seen in the output of any model.}
\label{plotsC13O16}
\end{narrow}
\end{figure*}

\begin{figure*}[tbp]
\centering
\begin{narrow}{0in}{0in}
\begin{minipage}[c]{.44\linewidth}
	\includegraphics[height=50mm]{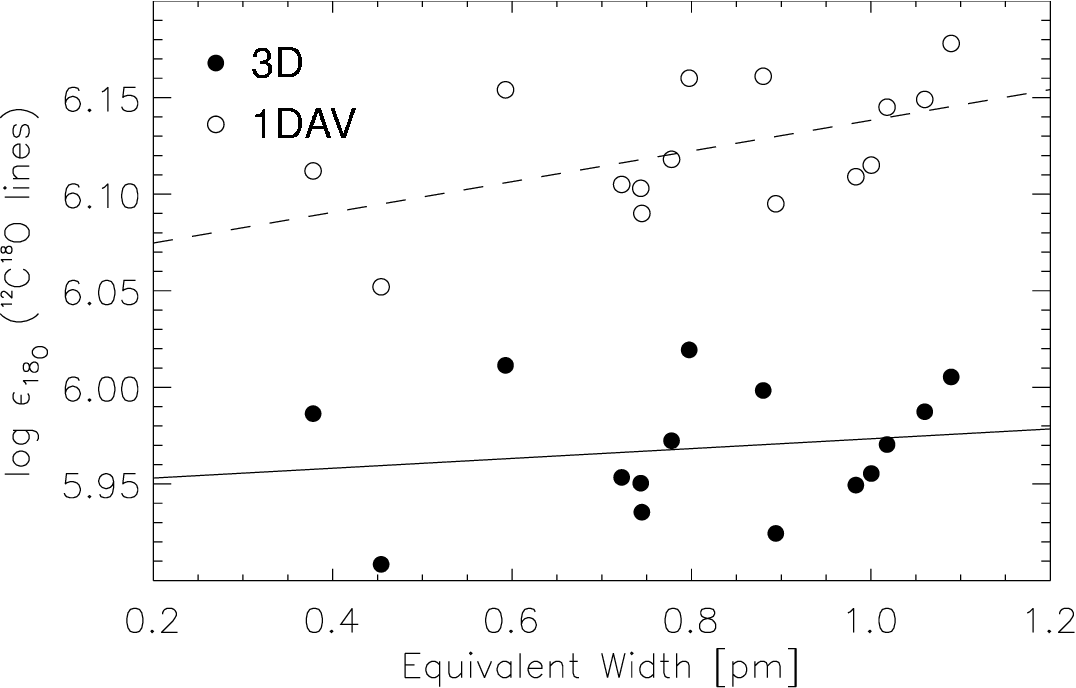}
	\\[5mm]
	\includegraphics[height=50mm]{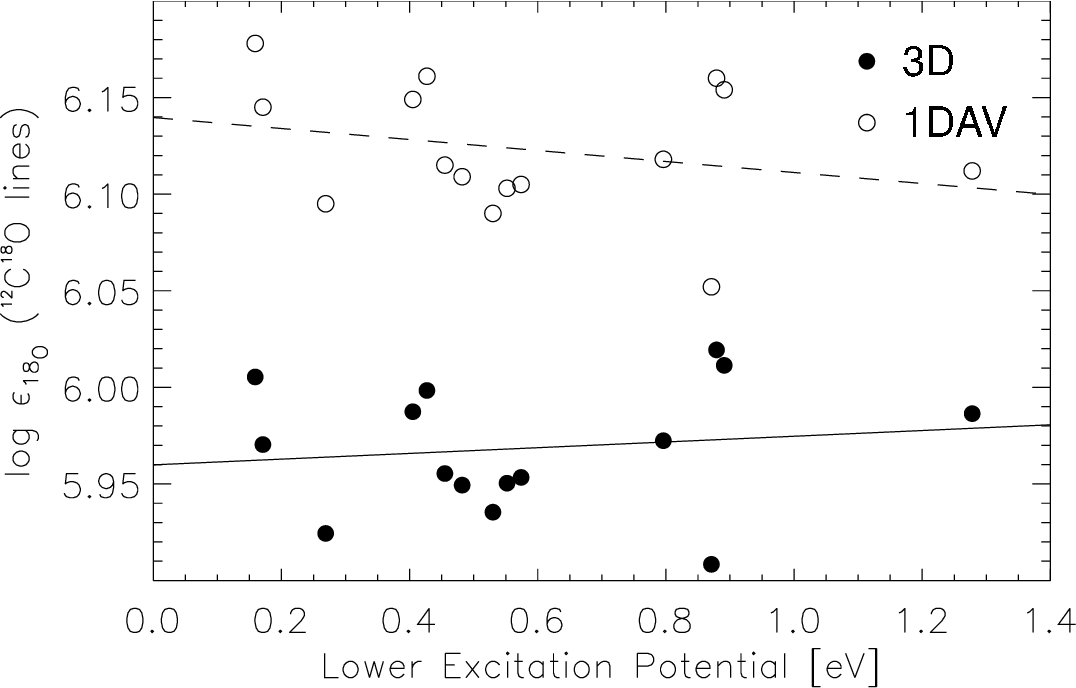}
\end{minipage}%
\hspace{0.12\linewidth}
\begin{minipage}[c]{.44\linewidth}
	\includegraphics[height=50mm]{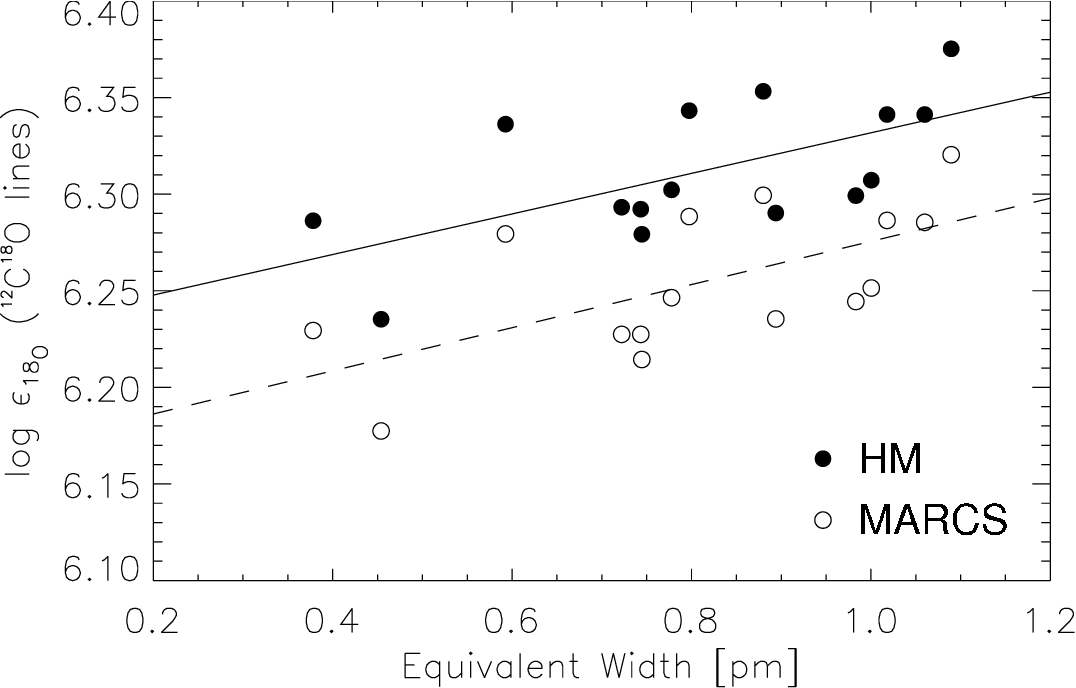}
	\\[5mm]
	\includegraphics[height=50mm]{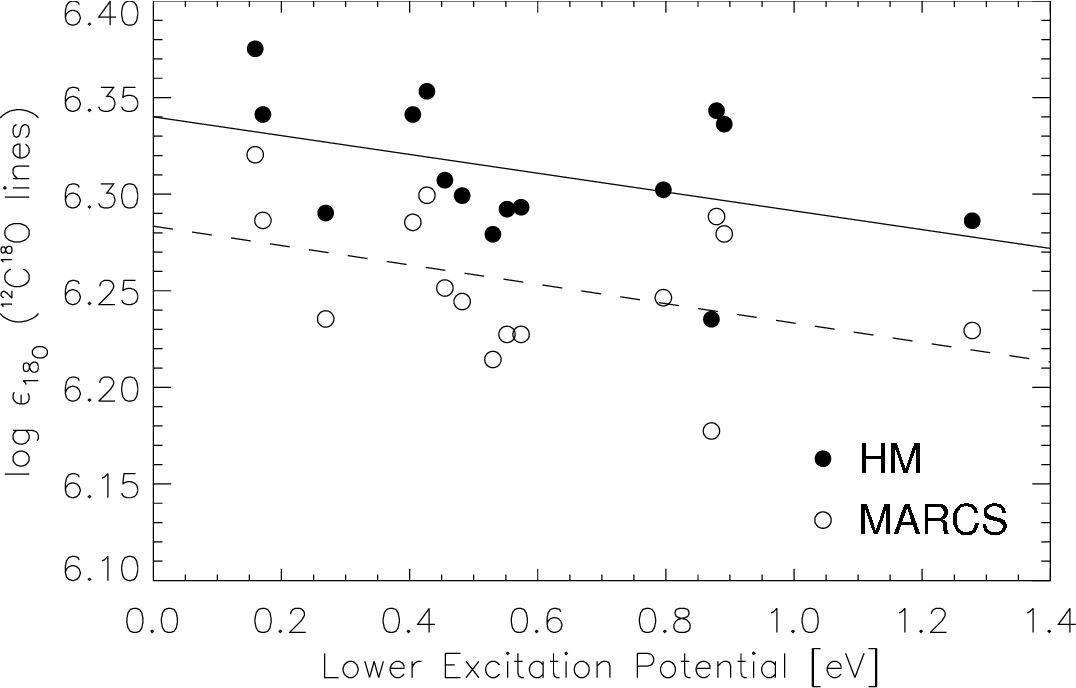}
\end{minipage}%
\caption[$^{18}$O abundance plots]{The same as Fig.~\protect\ref{plotsCOweak}, but indicating $^{18}$O through the use of the $^{12}$C$^{18}$O lines.  Significant trends can be seen in equivalent width and excitation potential in the HM and \textsc{marcs} results.}
\label{plotsC12O18}
\end{narrow}
\end{figure*}

\subsection{Results}
\label{isoresults}
Derived abundances are plotted as a function of equivalent width and excitation potential for each line in the weak $^{12}$C$^{16}$O (Fig.~\ref{plotsCOweak}), LE $^{12}$C$^{16}$O (Fig.~\ref{plotsCOLE}), \mbox{$\Delta v = 2$} $^{12}$C$^{16}$O (Fig.~\ref{plotsCOdv2}), $^{13}$C$^{16}$O (Fig.~\ref{plotsC13O16}) and $^{12}$C$^{18}$O sets (Fig.~\ref{plotsC12O18}).  The only appreciable trend evident in the 3D results is with equivalent width in the case of the LE lines, though these lines do also produce a smaller trend in excitation potential in 3D.  The HM and \textsc{marcs} models display trends in equivalent width and excitation potential for the $^{12}$C$^{18}$O lines, and excitation potential in the \mbox{$\Delta v = 2$} lists.  In addition, the \textsc{marcs} model exhibits a trend with equivalent width in its LE results.  In the case of the 1DAV model, no significant trends are seen except for a small slope with equivalent width in the $^{12}$C$^{18}$O list.

Final abundance measures and isotopic ratios as defined by the weak, LE and \mbox{$\Delta v = 2$} $^{12}$C$^{16}$O lines are tabulated (Table~\ref{tableCO}).  Errors in $^{12}$C abundances are given by single standard deviations within individual $^{12}$C$^{16}$O line lists.  Errors in bulk C abundances are single standard deviations within each $^{12}$C$^{16}$O list when a fractional scalefactor (cf. Appendix~\ref{fracsf}) was employed, as uncertainty in the scalefactor due to isotopic abundance uncertainties was negligible compared to line-to-line scatter.  Similarly, errors in isotopic abundances are a single standard deviation within the relevant isotopomeric list.  Errors in isotopic ratios were determined in logarithmic units as the sum in quadrature of errors attached to logarithmic $^{12}$C and isotopic abundances.  Asymmetric errors in absolute ratios are given as corresponding deviations due to these calculated (symmetric) logarithmic errors.  Overall results in 3D from different indicators are summarised in Table~\ref{summary}.

\subsection{The Solar C Abundance}
\label{cdiscuss}

The absence of significant trends in 3D model abundances over the primary (weak) and overtone (\mbox{$\Delta v = 2$}) line lists (Figs.~\ref{plotsCOweak} and \ref{plotsCOdv2}), as well as the similarity of the resulting average abundances (\mbox{$\log\epsilon_\mathrm{C}$ = 8.40, 8.37}), give us confidence in the accuracy of these values.  They also suggest that the 3D model is accurate in the quite high atmospheric layers in which these lines form.  The agreement of abundances derived using the normal weak and overtone lines is quite an achievement, as these indicators have not generally produced consistent results.  The HM model for example produces abundances differing by 0.08 dex between the two lists \citep[and by nearly three times this between CO lines and other indicators of carbon abundance stated in ][]{AspVI}.  It also exhibits a clear trend with excitation potential for the \mbox{$\Delta v = 2$} lines (bottom-right of Fig.~\ref{plotsCOdv2}), further suggesting the inadequacy of its description of overtone CO.  The same trends are true of \textsc{marcs}, though it at least produces consistent abundances between the two line lists.  That the 1DAV model abundances in these two cases are much closer to the 3D results, though still not quite as low, reflects the fact that both the mean temperature structure and the temperature inhomogeneities play a role in producing different CO-derived C abundances in 3D than with HM or \textsc{marcs}.

The same success was not seen in LE results, where a striking trend with equivalent width is present in the 3D case (top left of Fig.~\ref{plotsCOLE}).  The reason for this is very likely the larger range of formation depths of these lines and increasing inadequacies of the 3D model atmosphere in higher layers.  The weaker of the LE lines can be seen to produce abundance measures very close to the primary and overtone results, as they form at similar heights (cf. Fig.~\ref{formationheights}).  However, the derived abundance increases with greater equivalent width and line formation height, resulting in an average abundance quite a bit higher than the primary or overtone diagnostics.  

\begin{table*}[tbp]
\centering
\caption[3D Abundances and isotopic ratios implied by all $^{12}$C$^{16}$O lists]{Summary of carbon abundances and isotopic ratios produced with the 3D model by the three different $^{12}$C$^{16}$O line lists, as well as the adopted values.  Note the poor agreement of the LE results with the other two, reflecting the trend seen in Fig.~\protect\ref{plotsCOLE}.  Adopted values were calculated via 2:0:1 weightings of the \mbox{weak:LE:$\Delta v = 2$} lists.}
\label{summary}
\begin{tabular}{l c c c c c}
\hline
$^{12}$C$^{16}$O List & Weak & LE & $\Delta v = 2$ & Adopted & Terrestrial\\
\hline
$\log \epsilon_\mathrm{C}$ & $8.40\pm0.01$ & $8.48\pm0.04$ & $8.37\pm0.01$ & $8.39\pm0.05$ &\vspace{0.5mm}\\
$^{12}$C/$^{13}$C & 88.8$^{+5.3}_{-5.0}$ & 107.6$^{+13}_{-12}$ & 82.8$^{+5.2}_{-4.9}$ & 86.8$^{+3.9}_{-3.7}$ & $89.4\pm 0.2$ \vspace{1mm}\\
$^{16}$O/$^{18}$O & 490$^{+41}_{-37}$ & 594$^{+80}_{-71}$ & 457$^{+39}_{-36}$ & 479$^{+29}_{-28}$ & $498.7\pm 0.1$ \vspace{1mm}\\
\hline
\end{tabular}
\end{table*}

A number of reasons could be postulated for this discrepancy.  The failure of the chemical equilibrium approximation (ICE) with height in the atmosphere would mean that CO density was being overpredicted in the models, producing stronger lines than otherwise would result from a particular abundance.  Hence, less abundance would be required to reproduce a given line profile than in the non-ICE case, not more as is seen here.  The breakdown of LTE at height in the atmosphere could possibly cause overestimated abundances here, though it seems unlikely given the repeated conclusion that CO lines form in LTE \citep{AW89, UitII}.  Another possibility could be that temperature contrast in the upper layers of the model is too low, as an increase in this contrast would produce lower temperature cool regions, which contribute more to increasing line strength than hotter regions would to decrease it, due to the increase in CO and the nonlinear temperature dependence of line formation.  However, this seems unlikely as our bisector investigations hinted at too much temperature contrast high in the atmosphere.   

Also possible is that the velocity structure of the upper atmosphere is slightly incorrect in the 3D simulations.  The \textsc{marcs} model exhibited a similar equivalent width trend in implied abundances to the 3D model for the LE lines, also suggesting problems in its upper layers.  However, even the strongest LE lines were not sensitive to the microturbulence parameter used with the \textsc{marcs} model, making this explanation unlikely.  Similarly, other controls revealed that the problem was not due to the adopted collisional damping parameter.  The most likely explanation for the pronounced trend seen in the 3D LE results is that the mean temperature structure at height is slightly too high, causing a reduction in line strength for a given abundance and conversely, increased abundances for given line strengths.  This is consistent with the lack of trend but higher average abundance indicated by the 1DAV model, as the horizontal averaging would smear out the effects of overly hot localised regions (e.g. intergranular lanes) at height but produce higher abundances overall due to the complete absence of temperature inhomogeneities.  This is also in line with the bisector discrepancies seen using strong CO lines.

The regenerated model of Sect.~\ref{coresults} was trialled on the LE and weak $^{12}$C$^{16}$O lines to see if agreement could be improved, and also to check that the newer model did not produce significantly different abundances where the older model was thought to be accurate (i.e. the weak lines).  Derived abundances in both cases were approximately \mbox{0.03 dex} larger, virtually identical to the effect of low resolution upon iron abundances found by \citet{AspRes}.

Despite the poor performance of the 3D model for the LE lines, none of the 1D models managed to derive much better agreement between these and the weak $^{12}$C$^{16}$O lines (which are regarded as the best CO indicators of carbon abundance).  In the sense of the LE lines' disagreement with other diagnostics, the 3D model simply presented no improvement, rather than any loss in performance over HM and \textsc{marcs}.  In all three diagnostics, the carbon abundance in 3D is considerably lower than indicated by HM or \textsc{marcs}, as has generally been found for other species \citep[e.g.][]{AspII, AspIV, AspRev, AspVI}.  This is a general consequence of extending atmospheric simulations to three dimensions, due to the permission of temperature inhomogeneities and the nonlinear temperature dependence of line formation.  That all three sets of $^{12}$C$^{16}$O lines continued to display this effect is a positive comment on the accuracy and consistency of the 3D model, especially given the past difficulties with 1D analyses of CO lines \citep{GandS95}.

Given the relative performance of the different $^{12}$C$^{16}$O line lists and their previously recognised suitability for carbon abundance determination, for the final analysis weak lines were given a weighting of 2, \mbox{$\Delta v = 2$} lines a weighting of 1 and LE lines no weighting at all.  We note however that abundances from the weakest LE lines are in good agreement with those from the two preferred line types.  Hence, on the basis of the lines measured in this study, the final carbon abundance arrived at is
\begin{displaymath}
\log\epsilon_\mathrm{C} = 8.39\pm0.05.
\end{displaymath}
The stated error is designed to encompass unquantified systematic errors from e.g. atomic data, the model atmosphere and equivalent width measurements; these systematics almost certainly outweigh any statistical variability amongst abundances indicated by individual lines or lists.
    
\subsection{The Solar $^{12}$C / $^{13}$C and $^{16}$O / $^{18}$O Ratios}
\label{isodiscuss}

The abundances produced by the individual $^{13}$C$^{16}$O and $^{12}$C$^{18}$O lines in 3D (left panels in Figs.~\ref{plotsC13O16} and \ref{plotsC12O18}) show no significant equivalent width or excitation potential dependence, hence indicating no major inadequacies in the relevant atmospheric layers of the model.  This is also the case for 1D calculations with the $^{13}$C$^{16}$O lines (right panels in Fig.~\ref{plotsC13O16}) but not the $^{12}$C$^{18}$O lines, where trends with equivalent width and excitation potential are evident in the HM and \textsc{marcs} results (right panels in Fig.~\ref{plotsC12O18}).  The two trends are explicitly linked, as the higher excitation lines form lower and therefore have smaller equivalent widths, so it is no surprise that one trend exists in equivalent width and the opposite is seen in excitation potential.  These trends would seem to indicate the superiority of the 3D model in the description of the very weak $^{12}$C$^{18}$O lines.

The derived isotopic ratios in Table~\ref{tableCO} using the different model atmospheres show a very strong separation between 3D and 1D results.  This is surprising at first, though suggests that in this context, the move to three dimensions is an important improvement.  This is highlighted in particular by the general difference between 3D and 1DAV results.  Our investigations have revealed that, due their lower excitation potentials, the isotopomeric lines are actually over 20\% more temperature sensitive than the weak $^{12}$C$^{16}$O lines.  At least in the case of the weak $^{12}$C$^{16}$O lines, this is almost certainly the reason for the greater decrease in isotopic abundances than the carbon abundance in 3D.  This is because the heterogeneic temperature structure of the 3D simulations (which is responsible for reduced abundances normally) has a greater decreasing effect upon the isotopomeric lines and therefore $^{13}$C and $^{18}$O abundances than it does upon the derived carbon abundance.  Because of this differential temperature sensitivity, the horizontal averaging implicit in 1D models means that they have difficulty reproducing the correct solar isotopic ratios.  Given their excitation potentials, the temperature sensitivities of the LE and \mbox{$\Delta v = 2$} lines are rather similar to the isotopomeric lines.  This is the reason for the better agreement between 1D and 3D isotopic ratios from the \mbox{$\Delta v = 2$} lines than from the weak $^{12}$C$^{16}$O lines.  Any similar effect in ratios derived from LE lines is lost in the distortion of the 3D carbon abundance by the observed trend with formation height.

Because each $^{12}$C$^{16}$O list produced different abundances and each was compared to the same $^{13}$C$^{16}$O and $^{12}$C$^{18}$O results, different 3D isotopic ratios (and absolute $^{18}$O abundances, since $\epsilon_\mathrm{O}$ was fixed) were obtained.  However, the LE ratios can be discarded due to the already established invalidity of the LE carbon abundance.  In the derivation of the final isotopic ratios, the \mbox{$\Delta v = 2$} ratios were given a lower weighting in the same manner as in the calculation of the final carbon abundance (i.e. half the weighting afforded the weak ratios).  Hence, the adopted isotopic ratios are:
\begin{align*}
^{12}\mathrm{C}/^{13}\mathrm{C} &= 86.8^{+3.9}_{-3.7}\\
^{16}\mathrm{O}/^{18}\mathrm{O} &= 479^{+29}_{-28}
\end{align*}
Errors are combinations of statistical errors in the weak and \mbox{$\Delta v = 2$} results.  We assume that the systematic errors discussed as relevant to the carbon abundance in Sect.~\ref{cdiscuss} effect all lines considered for ratios equally, so need not be included in final ratio errors.  This assumption is based on fact that the isotopic, weak and \mbox{$\Delta v = 2$} lines all form at approximately the same photospheric heights, reflecting essentially the same line formation processes.

\section{Comparisons with Previous Work}
\label{comparisondiscuss}

This study represents a significant step forward from the only other 3D investigation of CO line formation to date \citep{UitI}.  Firstly, we used a more modern 3D hydrodynamic model atmosphere code, whereas the previous study was based upon the very early \citet{SandN89} version, running at a resolution of just \mbox{63 $\times$ 63 $\times$ 63}.  Secondly, the current study produced profiles spatially averaged over the entire simulation domain and temporally averaged over about 100 snapshots corresponding to approximately \mbox{50 min} of solar time, whereas that of \citet{UitI} used a vertical slice through a single snapshot.  Thirdly, the maximum vertical extent of the current study was about \mbox{1.2 Mm}, compared with \mbox{0.6 Mm} in the earlier work.  Finally, agreement between the modelled spectrum and observation was excellent overall, with reasonable agreement even found at the level of individual line bisectors, a measure highly sensitive to small deviations of the simulation from reality.  As suggested by \citet{AR2003}, the overly cool version of the 3D model atmosphere employed by \citeauthor{UitI} is probably somewhat responsible for the overly deep CO line cores he found.  The recently reduced solar carbon and oxygen abundances \citep{AspVI, AspIV} have also played a role in the success of the current study, permitting the deviation of line profiles from the \citet{GandS98} abundances in order to improve agreement with observation.  This option was not realistically available to \citet{UitI} given prevailing wisdom at the time, and in light of the current results this probably also contributed to his modelled overprediction of CO line depths.  

In the context of spectral line formation, our results also constitute an improvement over the more recent 2D work of \citet{2D}.  This is due not only to the better agreement achieved with observation, but also the superiority of our model atmosphere in this context.  Apart from being 3D rather than 2D and therefore inherently more realistic, the model we employ includes line blanketing, rather than treating radiative transfer in the hydrodynamic code as grey.  In addition, our model exhibits an effective temperature in excellent agreement with the Sun's \citep[see][]{AspI}, whereas the snapshots \citeauthor{2D} employ have an effective temperature approximately 380K lower.  Whilst \citeauthor{2D} utilise non-equilibrium chemistry, they find that it is not required at the heights of formation considered in this paper, justifying our use of the ICE approximation throughout.

The adopted bulk carbon abundance is in excellent agreement with the \mbox{$\log\epsilon_\mathrm{C} = 8.39\pm 0.05$} of \citet{AspVI}, which was based on very different indicators (C\,\textsc{i}, [C\,\textsc{i}], C$_2$ and CH).  The current figure therefore firms our belief in the accuracy of both carbon results and the 3D model atmosphere.  In comparison, CO results with the preferred lines using the HM and \textsc{marcs} models do not agree at all with their C\,\textsc{i}, [C\,\textsc{i}], C$_2$ and CH counterparts given by \citet{AspVI}; CO lines give $\log\epsilon_\mathrm{C} = 8.60$, 8.69 but other lines 8.39--8.53 with HM, whilst with \textsc{marcs} CO lines indicate $\log\epsilon_\mathrm{C} = 8.55$, 8.58 and other lines give 8.35--8.46.  Given the height of formation of the CO lines, this result is a remarkable success for the 3D model.  The new carbon abundance constitutes a major reduction from the commonly adopted \citet{GandS98} value of \mbox{$\log\epsilon_\mathrm{C} = 8.52\pm 0.06$}, not to mention a multitude of even higher earlier estimates \citep[e.g.][]{HLG87, AG89}.

Using the HM model, \citet{HLG87} found isotopic ratios of \mbox{$^{12}$C/$^{13}$C = $84\pm5$} and \mbox{$^{16}$O/$^{18}$O = $440\pm50$}, broadly consistent with the ratios we obtain according to the stated errors.  In comparison, our own HM results (weighted amongst the lists in the same manner as the 3D ratios) were \mbox{$^{12}$C/$^{13}$C = 73.4$^{+2.8}_{-2.7}$} and \mbox{$^{16}$O/$^{18}$O = 368$^{+25}_{-23}$}.  The difference in HM results between the two studies evidently reflects some combination of the improved observations, line lists and molecular data used in the present work.\footnote{The $\log gf$ values were typically 0.02 dex larger in the \citet{HLG87} study, so these are unlikely to have affected ratios.  None of the weak $^{12}$C$^{16}$O lines we have used were employed by \citeauthor{HLG87}.  Equivalent widths for the few common isotopomeric lines were larger on average in their study, leading to an underprediction of isotopic abundances and therefore an overprediction of ratios.}  However, the effect of these improvements is to reduce the ratios, whereas the adopted ratios calculated in 3D are slightly greater than those previously found.  Hence, the agreement between our results and those of \citet{HLG87} is somewhat fortuitous; any difference between old and new \emph{adopted} ratios primarily reflects the shift from 1D to 3D modelling, with the effects of this shift in fact mostly nullified by improved input data.

We also believe that our ratios are more reliable than those very recently produced independently by \citeauthor{Ayres05} (\citeyear{Ayres05}; \mbox{$^{12}$C/$^{13}$C = $80\pm3$} and \mbox{$^{16}$O/$^{18}$O = $440\pm20$}, where errors are standard deviations rather than the lower standard error \citeauthor{Ayres05} favour).  They utilise our 1DAV model and a multicomponent version of it meant to approximate the true 3D case.  Discrepancies between isotopic ratios derived in the two studies using the 1DAV model can be explained by differences in measured equivalent widths and our use of the opacity scalefactors detailed in Sect.~\ref{opacitysf}.\footnote{Adopting \citeauthor{Ayres05}'s smaller equivalent widths, and considering the resultant average change in abundances over the small number of lines common to the two studies, as well as discarding the opacity scalefactors, we would obtain $^{12}$C/$^{13}\mathrm{C}=80.9$ and $^{16}$O/$^{18}\mathrm{O}=452$ for the 1DAV model using the weak $^{12}$C$^{16}$O list.  This is in comparison to $^{12}$C/$^{13}\mathrm{C}=80.7\pm3.8$ and $^{16}$O/$^{18}\mathrm{O}=443\pm31$ obtained by \citeauthor{Ayres05} using the same model.}.  We believe our equivalent width measurements to be superior: whilst \citet{Ayres05} utilise Gaussian fits to ATMOS data, we performed direct integration upon line profiles, since the ATMOS ATLAS-3 data is of sufficiently high signal-to-noise to make this possible and observed profiles always exhibit more extended wings than a pure Gaussian.  Furthermore, all \citeauthor{Ayres05}'s Gaussian fits to isotopomeric lines are constrained to have a FWHM of \mbox{4.05 kms$^{-1}$}, whereas our integration allowed this value to be directly determined.  However, \citeauthor{Ayres05} choose to discard the results of our model in favour of an empirical 1D model based upon a `Double-Dip' photospheric structure, which has no corroborating physical or theoretical basis beyond its ability to match the observed centre-to-limb variation of CO lines and the solar continuum.  The derived abundances show marked trends in excitation potential and a disagreement between fundamental and overtone bands, which our model does not (either in this paper or that of \citealt{Ayres05}).  To compensate for these trends, the authors introduce a highly-questionable, excitation potential-dependent change in given $gf$ values.  \citeauthor{Ayres05}'s choice of model is made upon the basis of the 1DAV model failing to properly reproduce the observed continuum and CO core centre-to-limb behaviour.  However, the stated discrepancy in the continuum centre-to-limb results is about double that produced by our own calculations.  We intend to return to this important issue in a subsequent study, where the temperature structure and full results of a large number of observational tests will be described.

The $^{12}$C/$^{13}$C ratio we find is in excellent agreement with the \citet{IUPAC02} terrestrial value of $89.4\pm0.2$.  Our ratio corresponds to a $\delta^{13}$C value of $30^{+46}_{-44}$ (without taking into account the uncertainty in the $\delta$-scale), in agreement with some of the lunar regolith measures discussed by \citet{WBBW}.  Our ratio does not agree with the $\delta^{13}\mathrm{C}\leq -105\pm20$ measured by \citet{hash04}, suggesting that theories for the formation of organics in the early solar system based upon this low value may require revision.  We confirm a high solar $^{12}$C/$^{13}$C ratio relative to the local ISM ($62\pm4$: \citealt{LangPenz}; $68\pm15$: \citealt{savage05}).  This may support the notion of a galactic enrichment of $^{13}$C over the Sun's lifetime \citep[e.g.][]{savage05}.  Alternatively, the difference could be due to greater isotopic inhomogeneity in the galaxy than previously thought, and migration of the solar system through it.

Our derived $^{16}$O/$^{18}$O abundance is also consistent with the terrestrial ratio of $498.7\pm 0.1$ \citep{IUPAC02}.  The corresponding $\delta^{18}$O value is $41^{+67}_{-59}$ (with error in the $\delta$-scale not accounted for).  Despite the lack of error bars on the $\delta^{18}\mathrm{O} = -50$ prediction of the self-shielding model \citep{YK04, Yin}, it seems unlikely that our results would be consistent with this prediction.  However, our result is in excellent agreement with the $\delta^{18}\mathrm{O} = 50$ recently determined by \citet{Ireland06} from lunar grains irradiated by the solar wind.  However, given uncertainty about how well the solar wind actually reflects the solar composition, and the size of our estimated errors, this amazing agreement could be slightly fortuitous.

\section{Conclusions}
\label{conclusions}

This is the most successful modelling of CO line formation in the solar atmosphere to date.  Solar abundance determinations using CO lines produced a carbon abundance of \mbox{$\log\epsilon_\mathrm{C} = 8.39\pm 0.05$}, and isotopic ratios of \mbox{$^{12}$C/$^{13}$C = 86.8$^{+3.9}_{-3.7}$} and \mbox{$^{16}$O/$^{18}$O = 479$^{+29}_{-28}$}.  These results represent a significant improvement over those of the past due to the combination of a state-of-the-art convective 3D model atmosphere, updated atomic data, better line lists and more accurate observations.  The carbon abundance is in excellent agreement with the recent findings based upon entirely different indicators of \citet{AspVI}, suggesting that the past problems with CO-derived abundances have been solved.  Both ratios are in excellent agreement with corresponding telluric values, though the oxygen ratio is in even closer agreement with a non-terrestrial value inferred by the latest lunar regolith analysis \citep{Ireland06}.  Our results also support the existence of a COmosphere, with a representative temperature below 4000K.
 
\begin{acknowledgements}
We dedicate this article to the memory of Hartmut Holweger, who passed away in early April, 2006.  He made many important contributions to solar (and stellar) physics, including studies of the solar chemical composition and the atmospheric temperature structure, both topics of this work.  He will be deeply missed by his many colleagues and friends.

We are grateful to Regner Trampedach for assistance with computing 3D solar models and useful discussions, Mike Gunson for ATMOS advice and Charles Jenkins for statistical assistance.  Thanks to Tom Ayres for helpful discussions and providing the preprint of his latest paper, and to Frank Briggs and Nathan Kilah for further discussions.  PS acknowledges support from the ANU National Undergraduate Scholarship, MA from the Australian Research Council and NG from the Royal Belgian Observatory.  This work has made extensive use of the Astrophysical Data System.
\end{acknowledgements}

\appendix

\section{ATMOS Apodization}
\label{ATMOSappendix}
Knowing the instrument's resolving power, it becomes possible to simulate the observational effect of an FTS.  This is simply done by convolving modelled spectra with an instrumental line shape (ILS) characterised by the spectral resolution in the appropriate scale.

\begin{figure}[tbp]
	\centering
	\includegraphics[width=0.44\textwidth]{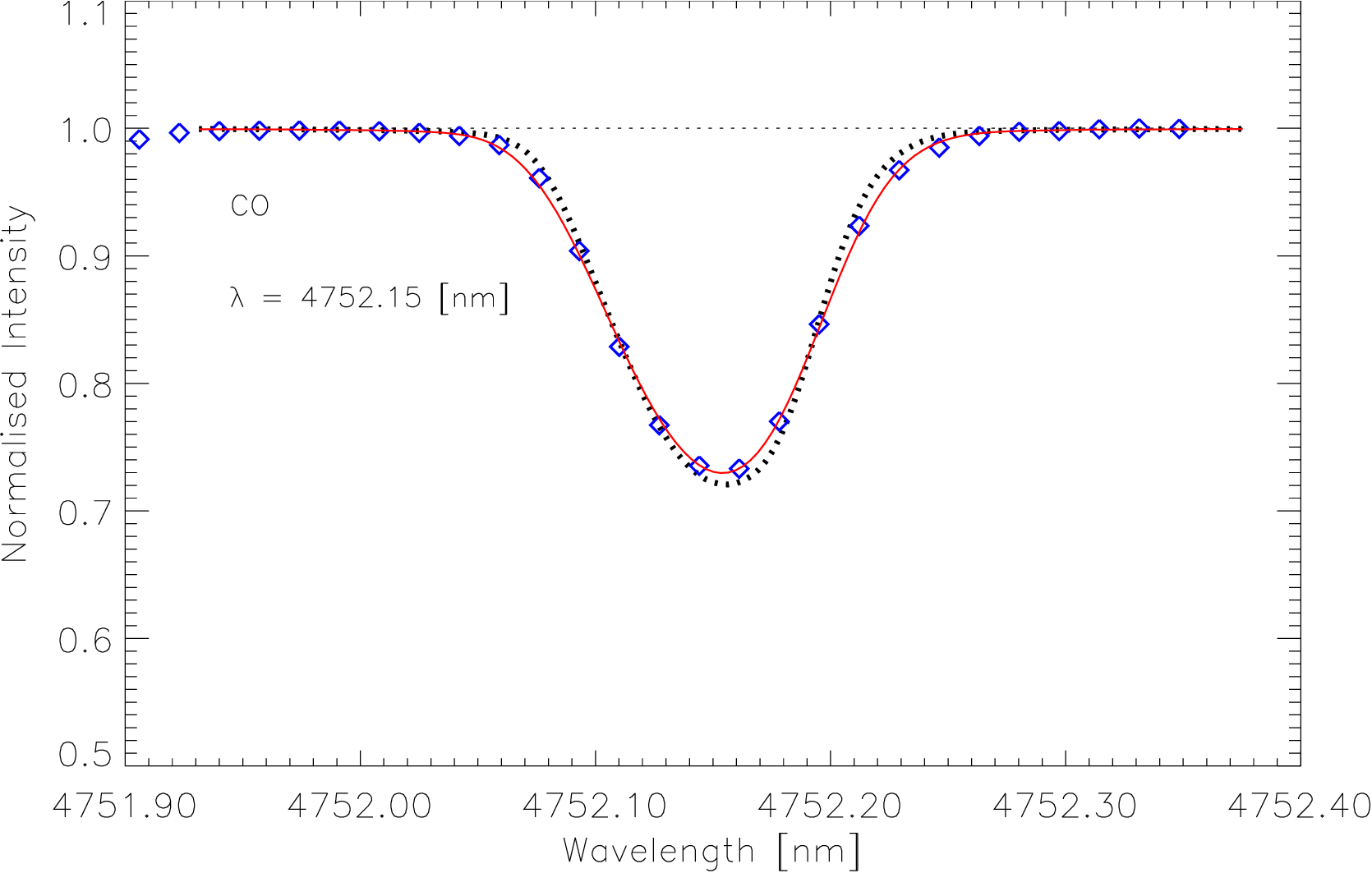}
\caption[Effects of medium BNA on \mbox{4752.2 nm} CO line]{The effects of medium BNA upon the same \mbox{4752.2 nm} CO line as shown in Fig.~\protect\ref{bisectors}.  The dotted profile has had no instrumental profile applied to it, whereas the solid line illustrates the effects of medium BNA characterised by \mbox{$\Delta\sigma$ = 1.5 kms$^{-1}$}.  Convolution with a simple sinc function of \mbox{$\Delta\sigma$ = 1.5 kms$^{-1}$} leaves the line profile virtually identical to its unconvolved counterpart.  In the interests of clarity this curve is not shown.  Note the far better agreement with the ATMOS spectrum of the apodized than the unapodized profile (reduced $\chi^2$ measures indicate an order of magnitude difference), vividly demonstrating the importance of correctly emulating instrumental effects when working with this data.} 
\label{bnaexample}
\end{figure}

Unfortunately however, FTS instruments are not always so well behaved as to exhibit perfect top-hat interferogram profiles.  This was certainly the case with the ATMOS instrument, which had an ILS that was far from a perfect sinc \citep{Gunson}.  To deal with the uncertainty in the true instrumental profiles of transform spectrometers, apodization is sometimes employed.  The process of apodization involves the multiplication of the interferogram with a profile other than the typical top-hat function, such as a Gaussian or simple triangle function, with somewhat tapered edges.  This of course has the same effect as convolving the spectral output with an ILS that is the Fourier transform of the apodizing function which one multiplies the interferogram by.  This smears out the variability in the instrumental profile, and dampens harmonic edge effects produced in the resultant spectrum by the finite maximum optical path difference of the FTS.  

The apodization applied by the ATMOS team \citep{ATMOS94, Gunson} was the medium function of \citet{BNA}:
{\setlength\arraycolsep{2pt}
\begin{eqnarray}
ILS(\sigma) & = & 0.26\mathrm{sinc}\,a - 0.464514\frac{\mathrm{sinc}\,a - \cos 
a}{a^2} - \nonumber\\
& & - 13.422570\frac{(1-3/a^2)\mathrm{sinc}\,a + (3/a^2)\cos a}{a^2},
\end{eqnarray}}where \mbox{$a = \frac{\pi\sigma}{\Delta\sigma}$}, $\sigma$ is the spectral variable and $\Delta\sigma$ its resolution.  This function is one of three purposefully created by \citeauthor{BNA} for their minimal resolutional broadening of spectral points and maximal damping of secondary maxima.  In order to correctly simulate the effects of the ATMOS instrument and the post-processing performed by its operators, we convolved modelled profiles with the same medium Beer-Norton function.  Model spectra were expressed in terms of Doppler velocity, so the convolving function argument was velocity resolution, i.e. \mbox{$\frac{c}{R} = $ 1.5~km\,s$^{-1}$}.  The striking effect of this post-processing is seen in Fig.~\ref{bnaexample}, where a predicted 3D line profile that has undergone Beer-Norton apodization (BNA) fits ATMOS data far better than either its unconvolved counterpart, or a profile convolved with a plain sinc function.

\section{Scalefactors}
\label{scalefactors}

\subsection{Mass scalefactor}
\label{masssf}
The mass scalefactor allowed the correct molecular mass to be used in the line formation calculations, and was simply the ratio of the mass of the isotopomer in question to that of the most common isotopomer (i.e. $^{12}$C$^{16}$O).  These factors, 1.03583 for $^{13}$C$^{16}$O and 1.07157 for $^{12}$C$^{18}$O, were calculated from nuclear weights given by \citet{weights}.  

\subsection{Opacity Scalefactor}
\label{opacitysf}
The opacity scalefactor contained the information about different isotopomer concentrations and therefore the atomic isotopic ratios, as the line opacities for any particular transition in two different isotopomers are proportional to their densities.  The use of an opacity scalefactor was preferable to simply altering the input carbon abundance in order to emulate the difference in isotopic abundances.  This is because the overall carbon abundance was not altered and therefore did not feed back on other parts of the simulation like CN line formation or the temperature structure, which might in turn indirectly effect CO line formation.  The opacity scalefactor was necessary anyway, in order to include an actual molecular difference (rather than just that due to abundances) in opacity between isotopomers.  This opacity correction arises as follows:

First, recall that Doppler (thermal) line broadening by a species of mass $m$ has a narrow Gaussian characteristic profile of the form
\begin{equation}
\label{eq8}
\phi_\mathrm{D}(\Delta\lambda) = \frac{1}{\sqrt{\pi}\Delta\lambda_\mathrm{D}}e^{-(\Delta\lambda/\Delta\lambda_\mathrm{D})^2},
\end{equation}
where $\Delta\lambda$ is the distance from the line centre of wavelength $\lambda_0$ and
\begin{equation}
\label{eq9}
\Delta\lambda_\mathrm{D} = \frac{\lambda_0}{c} \sqrt{\frac{2kT}{m} + \xi_\mathrm{t}^2}
\end{equation}
is the Doppler width of the Gaussian.  The microturbulent velocity $\xi_\mathrm{t}$ is an extra `fudge factor' used in analyses based on 1D model atmospheres, introduced to emulate Doppler broadening due to inherently three-dimensional microturbulence in the gas.

Now, consider two isotopomers of a diatomic molecule of elements X and Y: `$\mathcal{A}$' = $^a$X$^c$Y and `$\mathcal{B}$' = $^b$X$^c$Y.  The opacity of a species in a given transition depends on the density of absorbers, the transition probability and the relevant line broadening effects.  In the case of a weak lines, as those considered in this study are, Doppler broadening given by the profile $\phi_\mathrm{D}$ from Eq.~\ref{eq8} can be approximated as the only significant contributor to the latter, meaning that for our molecule $\mathcal{B}$,
\begin{equation}
\label{kappa1}
\kappa_\mathcal{B^\star} \varpropto N(\mathcal{B}^\star)f_{\mathcal{B}^\star}\phi_\mathrm{D}
\end{equation}
where $\mathcal{B}^\star$ denotes the isotopomer $\mathcal{B}$ in some given energy level from which the transition in question occurs, $N(\mathcal{B}^\star)$ refers to the number density of isotopomers $\mathcal{B}$ in the excited energy level and $f_{\mathcal{B}^\star}$ refers to the oscillator strength or probability of the particular transition from the excited energy level.  Now, at any given wavelength width $\Delta\lambda$ in Eq.~\ref{eq8}, we see that
\begin{align}
\phi_\mathrm{D} &\varpropto \frac{1}{\Delta\lambda_\mathrm{D}}\hspace{2pt}\varpropto\hspace{2pt}\frac{1}{\sqrt{\frac{2kT}{m} + \xi_\mathrm{t}^2}}
\nonumber\\
\therefore \kappa_\mathcal{B^\star} &\varpropto \frac{N(\mathcal{B}^\star)f_{\mathcal{B}^\star}}{\sqrt{\frac{2kT}{m_\mathcal{B}} + \xi_\mathrm{t}^2}}.
\label{kappa2}
\end{align}

Now, assuming LTE, we know from statistical mechanics \citep{mihalasSA} that a Boltzmann distribution of energy levels implies that where many instances of some microscopic system $\Upsilon$ occur, the population of a particular excited energy level $\Upsilon^\star$ is given by
\begin{equation}
N(\Upsilon^\star) = \frac{N(\Upsilon)g_{\Upsilon^\star}\exp(\frac{-E_{\Upsilon^\star}}{kT})}{Z(\Upsilon)},
\end{equation}
where $g_{\Upsilon^\star}$ is the statistical weight of the excited energy level, $E_{\Upsilon^\star}$ is its energy and $Z(\Upsilon)$ is the partition function of the system.  Hence,
\begin{equation}
\kappa_\mathcal{B^\star} \varpropto 
\frac{N(\mathcal{B})g_{\mathcal{B}^\star}\exp(\frac{-E_{\mathcal{B}^\star}}{kT})f_{\mathcal{B}^\star}}{Z(\mathcal{B})\sqrt{\frac{2kT}{m_\mathcal{B}} + \xi_\mathrm{t}^2}}.
\label{kappa3}
\end{equation}
Now, the number of molecules is related to the number density of the molecule's constituent atoms through the well-known Guldberg-Waage law, so
\begin{equation}
N(\mathcal{B}) = \frac{N(^b\mathrm{X})N(^c\mathrm{Y})}{\varkappa(\mathcal{B})},
\label{kappa4}
\end{equation}
where $\varkappa(\mathcal{B})$ is the equilibrium constant of the formation-dissociation reaction of the isotopomer $\mathcal{B}$, given \citep{schadee64, tatum66} by
\begin{equation}
\varkappa(\mathcal{B}) \varpropto \frac{Z(^b\mathrm{X})Z(^c\mathrm{Y})\mu_\mathcal{B}^\frac{3}{2}\exp(\frac{-D_\mathcal{B}}{kT})}{Z(\mathcal{B})}
\label{kappa5}
\end{equation}
where $D_\mathcal{B}$ and $\mu_\mathcal{B}$ are the dissociation energy and reduced mass respectively of isotopomer $\mathcal{B}$.  Hence, combining Eqs.~\ref{kappa3}--\ref{kappa5} we now see that
{\allowdisplaybreaks
\begin{align}
\kappa_\mathcal{B^\star} &\varpropto 
\frac{N(^b\mathrm{X})N(^c\mathrm{Y})}{\varkappa(\mathcal{B})Z(\mathcal{B})}\frac{gf_{\mathcal{B}^\star}\exp(\frac{-E_{\mathcal{B}^\star}}{kT})}{\sqrt{\frac{2kT}{m_\mathcal{B}} + \xi_\mathrm{t}^2}} \nonumber\\
&\varpropto 
\frac{N(^b\mathrm{X})N(^c\mathrm{Y})Z(\mathcal{B})}{Z(\mathcal{B})Z(^b\mathrm{X})Z(^c\mathrm{Y})\mu_\mathcal{B}^\frac{3}{2}\exp(\frac{-D_\mathcal{B}}{kT})}\times \nonumber \\
&\hspace{6mm}\times
\frac{gf_{\mathcal{B}^\star}\exp(\frac{-E_{\mathcal{B}^\star}}{kT})}{\sqrt{\frac{2kT}{m_\mathcal{B}} + \xi_\mathrm{t}^2}} \nonumber \\
\therefore\hspace{1cm}\kappa_\mathcal{B^\star} &\varpropto \frac{N(^b\mathrm{X})N(^c\mathrm{Y})gf_{\mathcal{B}^\star}\exp(\frac{(D_\mathcal{B}-E_{\mathcal{B}^\star}}{kT})}{Z(^b\mathrm{X})Z(^c\mathrm{Y})\mu_\mathcal{B}^\frac{3}{2}\sqrt{\frac{2kT}{m_\mathcal{B}} + \xi_\mathrm{t}^2}},
\label{kappa6}
\end{align}
where $g_{\mathcal{B}^\star}$ and $f_{\mathcal{B}^\star}$ have been combined into $gf_{\mathcal{B}^\star}$, the $gf$ value of the line.}

So, the opacity scalefactor $\varsigma_\kappa$ for an isotopomer $\mathcal{B}$, where the reference isotopomer is $\mathcal{A}$ is then
\begin{align}
\varsigma_\kappa \equiv \frac{\kappa_\mathcal{B^\star}}{\kappa_\mathcal{A^\star}} &= \frac{\frac{N(^b\mathrm{X})N(^c\mathrm{Y})gf_{\mathcal{B}^\star}\exp(\frac{(D_\mathcal{B}-E_{\mathcal{B}^\star}}{kT})}{Z(^b\mathrm{X})Z(^c\mathrm{Y})\mu_\mathcal{B}^\frac{3}{2}\sqrt{\frac{2kT}{m_\mathcal{B}} + \xi_\mathrm{t}^2}}}
{\frac{N(^a\mathrm{X})N(^c\mathrm{Y})gf_{\mathcal{A}^\star}\exp(\frac{(D_\mathcal{A}-E_{\mathcal{A}^\star}}{kT})}{Z(^a\mathrm{X})Z(^c\mathrm{Y})\mu_\mathcal{A}^\frac{3}{2}\sqrt{\frac{2kT}{m_\mathcal{A}} + \xi_\mathrm{t}^2}}} \nonumber\\
&= \frac{N(^b\mathrm{X})gf_{\mathcal{B}^\star}Z(^a\mathrm{X})\mu_\mathcal{A}^\frac{3}{2}\exp(\frac{(D_\mathcal{B}-E_{\mathcal{B}^\star}}{kT})}
{N(^a\mathrm{X})gf_{\mathcal{A}^\star}Z(^b\mathrm{X})\mu_\mathcal{B}^\frac{3}{2}\exp(\frac{(D_\mathcal{A}-E_{\mathcal{A}^\star}}{kT})}\times \nonumber \\
&\hspace{6mm}\times
\frac{\sqrt{\frac{2kT}{m_\mathcal{A}} + \xi_\mathrm{t}^2}}{\sqrt{\frac{2kT}{m_\mathcal{B}} + \xi_\mathrm{t}^2}},
\label{kappa7}
\end{align}
Now, we make the approximations \mbox{$gf_{\mathcal{B}^\star} \approx gf_{\mathcal{A}^\star}$}, \mbox{$D_\mathcal{B} \approx D_\mathcal{A}$} and \mbox{$E_{\mathcal{B}^\star} \approx E_{\mathcal{A}^\star}$}, since $\mathcal{A}$ and $\mathcal{B}$ are very nearly the same molecule so their transitions and energy levels will be almost identical.  Further, assuming \mbox{$Z(^b\mathrm{X}) \approx Z(^a\mathrm{X})$} due to the first-order dependence of atomic orbitals upon charge (and not mass), we arrive at
\begin{equation}
\varsigma_\kappa = \frac{N(^b\mathrm{X})\mu_\mathcal{A}^\frac{3}{2}\sqrt{\frac{2kT}{m_\mathcal{A}} + \xi_\mathrm{t}^2}}{N(^a\mathrm{X})\mu_\mathcal{B}^\frac{3}{2}\sqrt{\frac{2kT}{m_\mathcal{B}} + \xi_\mathrm{t}^2}},
\label{kappa8}
\end{equation}
where the final scalefactor is independent of transition.  The reduced mass opacity corrections used for $^{13}$C$^{16}$O and $^{12}$C$^{18}$O were therefore 0.9346 and 0.9293 respectively, where nuclear masses were again sourced from \citet{weights}.  In practice, the ratio of the square root terms enters the opacity anyway through the mass scalefactor, and the microturbulent velocity is not used in 3D (i.e. \mbox{$\xi_\mathrm{t} = 0$} was used) so these can be disregarded also.  The total opacity scalefactor applied in each case was therefore the product of the appropriate one of the reduced mass correction factors and the ratio of reference ($^{12}$C or $^{16}$O) isotope to that involved in the line being modelled ($^{13}$C or $^{18}$O), i.e.
\begin{equation}
\varsigma_{\kappa} = \frac{N(^b\mathrm{X})}{N(^a\mathrm{X})}\left(\frac{\mu_\mathcal{A}}{\mu_\mathcal{B}}\right)^\frac{3}{2} \equiv \left(\frac{^b\mathrm{X}}{^a\mathrm{X}}\right)\left(\frac{\mu_\mathcal{A}}{\mu_\mathcal{B}}\right)^\frac{3}{2}
\end{equation}
Had the opacity scalefactor reflected abundance differences only, not intrinsic atomic (i.e. reduced mass) corrections, our adopted 3D-based ratios would have been \mbox{$^{12}$C/$^{13}$C = 92.9$^{+4.1}_{-3.9}$} and \mbox{$^{16}$O/$^{18}$O = 516$^{+32}_{-30}$}.

The opacity scalefactors were then iteratively altered in calculations of the $^{13}$C$^{16}$O and $^{12}$C$^{18}$O line lists to reflect different isotopic ratios, in the same manner as the bulk carbon abundance was iterated to determine the `$^{12}$C' abundance using the three sets of $^{12}$C$^{16}$O lines.  The same bulk carbon abundance as used in the final iteration of the weak $^{12}$C$^{16}$C calculation was used in each case, and the isotopic ratios after some iteration $i$ given by
\begin{equation}
\left(\frac{^a\mathrm{X}}{^b\mathrm{X}}\right)_{i} = \left(\frac{^a\mathrm{X}}{^b\mathrm{X}}\right)_{i-1}\frac{10^{\Delta ^{12}\mathrm{C}_\mathrm{final}}}{10^{\Delta \mathrm{iso}_{i}}}.
\end{equation}
Here $\Delta \mathrm{iso}_{i}$ and $\Delta ^{12}\mathrm{C}_\mathrm{final}$ are the abundance corrections produced by the $i$th isotopomer iteration and the final iteration of whichever $^{12}$C$^{16}$O list is used to indicate the $^{12}$C abundance, respectively.

\subsection{Fractional Scalefactor}
\label{fracsf}
Following opacity scalefactor convergence, a further iteration was also performed on each of the sets of $^{12}$C$^{16}$O lines using a fractional scalefactor.  This accounts for the fact that not all carbon in CO lines exists in $^{12}$C$^{16}$O.  This scalefactor was calculated from the just derived $^{12}$C/$^{13}$C and $^{16}$O/$^{18}$O ratios, such that
\begin{displaymath}
\varsigma_\mathrm{frac} = \frac{N(^{12}\mathrm{C}^{16}\mathrm{O})}{N(\mathrm{CO})},
\end{displaymath}
and since
\begin{equation*}
N(\mathrm{CO}) = N(^{12}\mathrm{C}^{16}\mathrm{O})+N(^{13}\mathrm{C}^{16}\mathrm{O})+N(^{12}\mathrm{C}^{18}\mathrm{O})+N(^{12}\mathrm{C}^{17}\mathrm{O})
\end{equation*}
we see that
\begin{align}
\varsigma_\mathrm{frac} &= \left(1 \hspace{2mm} + \hspace{2mm} \frac{N(^{13}\mathrm{C}^{16}\mathrm{O})}{N(^{12}\mathrm{C}^{16}\mathrm{O})} \hspace{2mm} + \hspace{2mm} \frac{N(^{12}\mathrm{C}^{18}\mathrm{O})}{N(^{12}\mathrm{C}^{16}\mathrm{O})} \hspace{2mm} + \hspace{2mm} \right. \nonumber \\
&\hspace{6mm} \left. + \hspace{2mm} \frac{N(^{12}\mathrm{C}^{17}\mathrm{O})}{N(^{12}\mathrm{C}^{16}\mathrm{O})}\right)^{-1}\nonumber\vspace{1mm}\\
&= (1 + {^{13}\mathrm{C}}/{^{12}\mathrm{C}} +  {^{18}\mathrm{O}}/{^{16}\mathrm{O}} + {^{17}\mathrm{O}}/{^{16}\mathrm{O}})^{-1}.
\label{fracsfeq}
\end{align}
Therefore, $\varsigma_\mathrm{frac}$ represents the fraction of the bulk carbon abundance actually indicated by $^{12}$C$^{16}$O lines.  Seeing as the $^{12}$C$^{17}$O lines are so weak in the ATMOS spectrum, accurate derivation of the $^{16}$O/$^{17}$O ratio would not have been possible in this study, so the value of $^{16}$O/$^{17}$O used in the calculation of fractional scalefactors was the terrestrial value of $\sim$2630 \citep{IUPAC02}.  Being so incredibly small, the contribution of any difference in the $^{17}$O abundance between Earth and the Sun would have had almost no effect upon the resultant scalefactor.  Using these fractional scalefactors, the final iterations of the $^{12}$C$^{16}$O line lists indicated bulk solar carbon abundances.

\section{Line Lists}
\label{lists}

Refer to Tables \ref{biseclines}--\ref{dv2lines}.

\begin{table}[htbp]
\centering
\caption[Strong $^{12}$C$^{16}$O line list]{Strong $^{12}$C$^{16}$O line list used for the Sect.~\protect\ref{coresults} study, consisting of 31 lines.  $\log gf$ data calculated from \protect\citet{GChack94}.}
\label{biseclines}
\begin{tabular}{l@{}c@{\hspace{2mm}}c@{\hspace{2mm}}c@{\hspace{2mm}}c@{\hspace{2mm}}c@{\hspace{2mm}}c}
		\hline
		\multicolumn{2}{c}{Trans.} & Wavelength & $\sigma$ & $\log gf$ & Exc. \\
		& & vac.~(nm) & (cm$^{-1}$) & & (eV) \\
		\hline
    1--0 P&6  & 4717.6910 & 2119.681 & $-$4.171 & 0.010 \\
    1--0 P&7  & 4726.7267 & 2115.629 & $-$4.104 & 0.013 \\
    1--0 R&31 & 4450.3204 & 2247.029 & $-$3.412 & 0.236 \\
    1--0 R&42 & 4398.4026 & 2273.553 & $-$3.277 & 0.428 \\
    1--0 R&58 & 4341.8489 & 2303.166 & $-$3.129 & 0.807 \\
    1--0 R&59 & 4339.0480 & 2304.653 & $-$3.121 & 0.834 \\
    1--0 R&60 & 4336.3331 & 2306.096 & $-$3.113 & 0.862 \\
    1--0 R&61 & 4333.7041 & 2307.495 & $-$3.106 & 0.891 \\
    2--1 P&2  & 4741.2775 & 2109.136 & $-$4.350 & 0.267 \\
    2--1 R&0  & 4715.7219 & 2120.566 & $-$4.648 & 0.266 \\
    2--1 R&18 & 4582.1879 & 2182.364 & $-$3.355 & 0.346 \\
    2--1 R&30 & 4510.9808 & 2216.813 & $-$3.134 & 0.485 \\
    2--1 R&36 & 4480.4509 & 2231.918 & $-$3.053 & 0.579 \\
    2--1 R&41 & 4457.5278 & 2243.396 & $-$2.996 & 0.670 \\
    3--2 R&14 & 4666.9654 & 2142.720 & $-$3.292 & 0.577 \\
    3--2 R&27 & 4584.3568 & 2181.331 & $-$3.011 & 0.705 \\
    4--3 R&15 & 4719.1825 & 2119.011 & $-$3.146 & 0.843 \\
    4--3 R&32 & 4614.8044 & 2166.939 & $-$2.818 & 1.031 \\
    4--3 R&38 & 4584.7513 & 2181.143 & $-$2.742 & 1.129 \\
    5--4 R&19 & 4752.1533 & 2104.309 & $-$2.955 & 1.130 \\
    5--4 R&28 & 4696.2029 & 2129.380 & $-$2.788 & 1.229 \\
    5--4 R&31 & 4679.3794 & 2137.036 & $-$2.745 & 1.270 \\
    5--4 R&34 & 4663.4526 & 2144.334 & $-$2.703 & 1.316 \\
    5--4 R&39 & 4638.8782 & 2155.694 & $-$2.642 & 1.400 \\
    5--4 R&40 & 4634.2564 & 2157.844 & $-$2.631 & 1.418 \\
    5--4 R&48 & 4600.7586 & 2173.555 & $-$2.550 & 1.579 \\
    6--5 R&31 & 4739.9862 & 2109.711 & $-$2.674 & 1.521 \\
    6--5 R&58 & 4627.2312 & 2161.120 & $-$2.394 & 2.066 \\
    6--5 R&61 & 4619.1075 & 2164.920 & $-$2.372 & 2.146 \\
    6--5 R&68 & 4603.5454 & 2172.239 & $-$2.322 & 2.347 \\
    7--6 R&49 & 4718.1039 & 2119.496 & $-$2.411 & 2.093 \\   
		\hline
\end{tabular}
\end{table}

\begin{table}[htbp]
\centering
\caption[Weak $^{12}$C$^{16}$O line list]{Weak $^{12}$C$^{16}$O line list used for the Sect.~\protect\ref{abunresults} study, consisting of 13 lines.  $\log gf$ data calculated from \protect\citet{GChack94}.}
\label{weaklines}
\begin{tabular}{l@{}c@{\hspace{2mm}}c@{\hspace{2mm}}c@{\hspace{2mm}}c@{\hspace{2mm}}c@{\hspace{2mm}}c@{\hspace{2mm}}c}
			\hline
			\multicolumn{2}{c}{Trans.} & Wavelength & $\sigma$ & $\log gf$ & Exc. & W$_\lambda$ & $\log\epsilon_{^{12}\mathrm{C}}$ \\
			& & vac.~(nm) & (cm$^{-1}$) & & (eV) & (pm) & (3D) \\
			\hline
    \multicolumn{2}{c}{1--0} & & & & & & \\ 
    P&96  & 6112.9978 & 1635.859 &$-$3.053 & 2.154 & 7.74 & 8.37 \\
    P&98  & 6162.3540 & 1622.757 &$-$3.046 & 2.242 & 6.72 & 8.38 \\
    R&98  & 4297.1649 & 2327.116 &$-$2.889 & 2.242 & 8.46 & 8.41 \\
    R&99  & 4297.8564 & 2326.741 &$-$2.883 & 2.286 & 7.72 & 8.39 \\
    R&100 & 4298.6396 & 2326.317 &$-$2.879 & 2.331 & 7.21 & 8.40 \\
    R&101 & 4299.5151 & 2325.844 &$-$2.873 & 2.376 & 6.69 & 8.39 \\
    R&103 & 4301.5437 & 2324.747 &$-$2.867 & 2.467 & 5.74 & 8.40 \\
    R&105 & 4303.9460 & 2323.449 &$-$2.857 & 2.560 & 4.95 & 8.40 \\
    R&107 & 4306.7253 & 2321.950 &$-$2.848 & 2.654 & 4.02 & 8.39 \\
    R&109 & 4309.8852 & 2320.247 &$-$2.842 & 2.750 & 3.38 & 8.40 \\
    \multicolumn{2}{c}{2--1} & & & & & & \\ 
    P&97  & 6225.7984 & 1606.220 &$-$2.755 & 2.443 & 7.67 & 8.39 \\
    P&98  & 6251.2034 & 1599.692 &$-$2.752 & 2.486 & 7.11 & 8.39 \\
    P&101 & 6329.1932 & 1579.980 &$-$2.742 & 2.619 & 5.61 & 8.38 \\
\hline
\end{tabular}
\end{table}

\begin{table}[htbp]
\centering
\caption[$^{13}$C$^{16}$O line list]{$^{13}$C$^{16}$O line list used for the Sect.~\protect\ref{abunresults} study, consisting of 16 lines.  $\log gf$ data calculated from \protect\citet{GChack94}.}
\label{1316}	
\begin{tabular}{l@{}c@{\hspace{2mm}}c@{\hspace{2mm}}c@{\hspace{2mm}}c@{\hspace{2mm}}c@{\hspace{2mm}}c@{\hspace{2mm}}c}
			\hline
			\multicolumn{2}{c}{Trans.} & Wavelength & $\sigma$ & $\log gf$ & Exc. & W$_\lambda$ & $\log\epsilon_{^{13}\mathrm{C}}$ \\
			& & vac.~(nm) & (cm$^{-1}$) & & (eV) & (pm) & (3D) \\
			\hline
     \multicolumn{2}{c}{1--0} & & & & & & \\
     R&42 & 4501.6474 & 2221.409 & $-$3.297 & 0.409 & 4.26 & 6.43 \\
     R&43 & 4497.3756 & 2223.519 & $-$3.286 & 0.429 & 4.25 & 6.44 \\
     R&44 & 4493.1896 & 2225.591 & $-$3.275 & 0.448 & 4.24 & 6.45 \\
     R&47 & 4481.1460 & 2231.572 & $-$3.245 & 0.511 & 4.02 & 6.46 \\
     R&55 & 4452.7709 & 2245.793 & $-$3.173 & 0.695 & 2.91 & 6.43 \\
     R&56 & 4449.6042 & 2247.391 & $-$3.165 & 0.720 & 2.77 & 6.42 \\
     R&54 & 4456.0218 & 2244.154 & $-$3.182 & 0.671 & 2.78 & 6.39 \\
     \multicolumn{2}{c}{2--1} & & & & & & \\
     R&46 & 4539.7468 & 2202.766 & $-$2.963 & 0.745 & 3.92 & 6.42 \\
     R&47 & 4535.7888 & 2204.688 & $-$2.951 & 0.766 & 4.11 & 6.46 \\
     R&49 & 4528.1342 & 2208.415 & $-$2.932 & 0.809 & 3.79 & 6.45 \\
     R&53 & 4513.8671 & 2215.395 & $-$2.896 & 0.901 & 3.46 & 6.47 \\
     R&62 & 4486.8117 & 2228.754 & $-$2.827 & 1.132 & 2.23 & 6.43 \\
     R&67 & 4474.7857 & 2234.744 & $-$2.790 & 1.275 & 1.80 & 6.45 \\
     R&77 & 4457.1692 & 2243.577 & $-$2.726 & 1.591 & 0.99 & 6.46 \\
     R&79 & 4454.6795 & 2244.830 & $-$2.714 & 1.660 & 0.89 & 6.48 \\
     \multicolumn{2}{c}{3--2} & & & & & & \\
     R&62 & 4542.4821 & 2201.440 & $-$2.658 & 1.380 & 1.88 & 6.46 \\
\hline
\end{tabular}
\end{table}

\begin{table}[htbp]
\centering
\caption[$^{12}$C$^{18}$O line list]{$^{12}$C$^{18}$O line list used for the Sect.~\protect\ref{abunresults} study, consisting of 15 lines.  $\log gf$ data calculated from \protect\citet{GChack94}.}
\label{1218}
\begin{tabular}{l@{}c@{\hspace{2mm}}c@{\hspace{2mm}}c@{\hspace{2mm}}c@{\hspace{2mm}}c@{\hspace{2mm}}c@{\hspace{2mm}}c}
			\hline
			\multicolumn{2}{c}{Trans.} & Wavelength & $\sigma$ & $\log gf$ & Exc. & W$_\lambda$ & $\log\epsilon_{^{18}\mathrm{O}}$ \\
			& & vac.~(nm) & (cm$^{-1}$) & & (eV) & (pm) & (3D) \\
			\hline
     \multicolumn{2}{c}{1--0} & & & & & & \\
     P&26 & 5033.4006 & 1986.728 & $-$3.572 & 0.159 & 1.09 & 6.01 \\
     P&27 & 5044.7281 & 1982.267 & $-$3.556 & 0.171 & 1.02 & 5.97 \\
     P&34 & 5127.5809 & 1950.237 & $-$3.462 & 0.269 & 0.89 & 5.92 \\
     P&43 & 5243.7159 & 1907.045 & $-$3.370 & 0.427 & 0.88 & 6.00 \\
     R&48 & 4486.1030 & 2229.106 & $-$3.237 & 0.530 & 0.74 & 5.94 \\
     R&49 & 4482.3365 & 2230.979 & $-$3.228 & 0.552 & 0.74 & 5.95 \\
     R&50 & 4478.6546 & 2232.813 & $-$3.219 & 0.574 & 0.72 & 5.95 \\
     \multicolumn{2}{c}{2--1} & & & & & & \\
     P&25 & 5084.5153 & 1966.756 & $-$3.293 & 0.405 & 1.06 & 5.99 \\
     P&29 & 5130.8774 & 1948.984 & $-$3.233 & 0.455 & 1.00 & 5.96 \\
     P&31 & 5154.8355 & 1939.926 & $-$3.205 & 0.482 & 0.98 & 5.95 \\
     \multicolumn{2}{c}{3--2} & & & & & & \\
     P&35 & 5270.2261 & 1897.452 & $-$2.987 & 0.796 & 0.78 & 5.97 \\
     P&40 & 5336.4145 & 1873.917 & $-$2.932 & 0.879 & 0.80 & 6.02 \\
     \multicolumn{2}{c}{4--3} & & & & & & \\
     P&21 & 5167.4916 & 1935.175 & $-$3.077 & 0.871 & 0.45 & 5.91 \\
     P&23 & 5190.1890 & 1926.712 & $-$3.039 & 0.891 & 0.59 & 6.01 \\
     \multicolumn{2}{c}{5--4} & & & & & & \\
     P&34 & 5393.2644 & 1854.165 & $-$2.788 & 1.278 & 0.38 & 5.99 \\
\hline
\end{tabular}
\end{table}

\begin{table}[htbp]
\centering
\caption[LE $^{12}$C$^{16}$O line list]{Low excitation (LE) $^{12}$C$^{16}$O line list used for the Sect.~\protect\ref{abunresults} study, consisting of 15 lines.  $\log gf$ data calculated from \protect\citet{GChack94}.}
\label{LElines}
\begin{tabular}{l@{}c@{\hspace{2mm}}c@{\hspace{2mm}}c@{\hspace{2mm}}c@{\hspace{2mm}}c@{\hspace{2mm}}c@{\hspace{2mm}}c}
			\hline
			\multicolumn{2}{c}{Trans.} & Wavelength & $\sigma$ & $\log gf$ & Exc. & W$_\lambda$ & $\log\epsilon_{^{12}\mathrm{C}}$ \\
			& & vac.~(nm) & (cm$^{-1}$) & & (eV) & (pm) & (3D) \\
			\hline
     \multicolumn{2}{c}{2--1} & & & & & & \\
     P&2 & 4741.2775 & 2109.136 & $-$4.350 & 0.267 & 20.79 & 8.55 \\
     R&0 & 4715.7219 & 2120.566 & $-$4.648 & 0.266 & 15.14 & 8.47 \\
     \multicolumn{2}{c}{3--2} & & & & & & \\
     P&2 & 4801.2476 & 2082.792 & $-$4.181 & 0.530 & 18.37 & 8.51 \\
     \multicolumn{2}{c}{4--3} & & & & & & \\
     P&2 & 4862.5943 & 2056.515 & $-$4.063 & 0.789 & 15.37 & 8.47 \\
     P&3 & 4871.6187 & 2052.706 & $-$3.886 & 0.790 & 18.37 & 8.49 \\
     P&4 & 4880.7591 & 2048.862 & $-$3.762 & 0.792 & 20.84 & 8.54 \\
     R&0 & 4836.2087 & 2067.735 & $-$4.362 & 0.787 & 10.53 & 8.43 \\
     R&1 & 4827.6411 & 2071.405 & $-$4.060 & 0.788 & 15.50 & 8.48 \\
     \multicolumn{2}{c}{5--4} & & & & & & \\
     P&1 & 4916.3051 & 2034.048 & $-$4.273 & 1.044 & 7.86 & 8.42 \\
     P&3 & 4934.5350 & 2026.533 & $-$3.799 & 1.046 & 14.68 & 8.44 \\
     P&4 & 4943.8275 & 2022.724 & $-$3.674 & 1.048 & 17.11 & 8.48 \\
     P&6 & 4962.7720 & 2015.003 & $-$3.499 & 1.053 & 20.05 & 8.50 \\
     R&0 & 4898.5440 & 2041.423 & $-$4.272 & 1.043 & 7.58 & 8.39 \\
     R&2 & 4881.2454 & 2048.658 & $-$3.793 & 1.045 & 14.87 & 8.45 \\
     R&4 & 4864.4047 & 2055.750 & $-$3.569 & 1.048 & 18.93 & 8.50 \\
    
\hline
\end{tabular}
\end{table}

\begin{table*}[htbp]
\centering
\caption[\mbox{$\Delta v = 2$} $^{12}$C$^{16}$O line list]{First overtone (\mbox{$\Delta v = 2$}) $^{12}$C$^{16}$O line list used for the Sect.~\protect\ref{abunresults} study, consisting of 66 lines.  $\log gf$ data calculated from \protect\citet{GChack94}.}
\label{dv2lines}
\begin{narrow}{0in}{0in}
\begin{minipage}[c]{.5\linewidth}
	\centering
	\begin{tabular}{l@{}c@{\hspace{2mm}}c@{\hspace{2mm}}c@{\hspace{2mm}}c@{\hspace{2mm}}c@{\hspace{2mm}}c@{\hspace{2mm}}c}
			\hline
			\multicolumn{2}{c}{Trans.} & Wavelength & $\sigma$ & $\log gf$ & Exc. & W$_\lambda$ & $\log\epsilon_{^{12}\mathrm{C}}$ \\
			& & vac.~(nm) & (cm$^{-1}$) & & (eV) & (pm) & (3D) \\
			\hline
      \multicolumn{2}{c}{2--0} & & & & & & \\
      P&7  & 2363.1249 & 4231.685 & $-$6.223 & 0.013 & 1.04 & 8.34 \\
      P&9  & 2368.0105 & 4222.954 & $-$6.119 & 0.021 & 1.31 & 8.35 \\
      P&13 & 2378.3082 & 4204.670 & $-$5.971 & 0.043 & 1.73 & 8.36 \\
      P&20 & 2398.0498 & 4170.055 & $-$5.799 & 0.100 & 2.13 & 8.35 \\
      P&21 & 2401.0522 & 4164.841 & $-$5.780 & 0.110 & 2.22 & 8.36 \\
      P&22 & 2404.1007 & 4159.560 & $-$5.762 & 0.120 & 2.23 & 8.35 \\
      P&32 & 2437.1831 & 4103.098 & $-$5.622 & 0.251 & 2.22 & 8.35 \\
      P&38 & 2459.3740 & 4066.075 & $-$5.559 & 0.352 & 2.00 & 8.35 \\
      P&41 & 2471.1542 & 4046.692 & $-$5.532 & 0.408 & 1.85 & 8.34 \\
      P&42 & 2475.1846 & 4040.103 & $-$5.523 & 0.428 & 1.84 & 8.35 \\
      P&46 & 2491.8344 & 4013.108 & $-$5.491 & 0.512 & 1.61 & 8.35 \\
      P&53 & 2523.0648 & 3963.434 & $-$5.443 & 0.676 & 1.24 & 8.36 \\
      P&54 & 2527.7505 & 3956.087 & $-$5.435 & 0.701 & 1.13 & 8.33 \\
      R&9  & 2328.4851 & 4294.638 & $-$6.024 & 0.021 & 1.63 & 8.36 \\
      R&13 & 2322.0848 & 4306.475 & $-$5.867 & 0.043 & 2.11 & 8.36 \\
      R&16 & 2317.7138 & 4314.597 & $-$5.775 & 0.065 & 2.47 & 8.37 \\
      R&19 & 2313.7084 & 4322.066 & $-$5.695 & 0.090 & 2.68 & 8.36 \\
      R&22 & 2310.0671 & 4328.879 & $-$5.625 & 0.120 & 2.91 & 8.36 \\
      R&23 & 2308.9340 & 4331.003 & $-$5.606 & 0.131 & 2.94 & 8.36 \\
      R&25 & 2306.7886 & 4335.031 & $-$5.564 & 0.155 & 3.00 & 8.36 \\
      R&26 & 2305.7763 & 4336.934 & $-$5.545 & 0.167 & 3.08 & 8.37 \\
      R&41 & 2295.4058 & 4356.528 & $-$5.301 & 0.408 & 2.86 & 8.36 \\
      R&65 & 2297.7720 & 4352.042 & $-$5.041 & 1.009 & 1.22 & 8.34 \\
      R&66 & 2298.3892 & 4350.873 & $-$5.032 & 1.039 & 1.25 & 8.37 \\
      \multicolumn{2}{c}{3--1} & & & & & & \\
      P&13 & 2408.3440 & 4152.231 & $-$5.487 & 0.309 & 2.61 & 8.36 \\
      P&15 & 2413.8559 & 4142.749 & $-$5.430 & 0.322 & 2.83 & 8.36 \\
      P&16 & 2416.6808 & 4137.907 & $-$5.403 & 0.330 & 2.93 & 8.36 \\
      P&18 & 2422.4694 & 4128.019 & $-$5.357 & 0.346 & 3.13 & 8.37 \\
      P&22 & 2434.6085 & 4107.437 & $-$5.279 & 0.385 & 3.30 & 8.36 \\
      P&23 & 2437.7618 & 4102.124 & $-$5.262 & 0.396 & 3.30 & 8.35 \\
      P&29 & 2457.6960 & 4068.851 & $-$5.174 & 0.471 & 3.38 & 8.36 \\
      P&30 & 2461.1901 & 4063.075 & $-$5.162 & 0.485 & 3.42 & 8.36 \\
	\hline
	\end{tabular}
\end{minipage}%
\begin{minipage}[c]{.5\linewidth}
	\centering
	\begin{tabular}{l@{}c@{\hspace{2mm}}c@{\hspace{2mm}}c@{\hspace{2mm}}c@{\hspace{2mm}}c@{\hspace{2mm}}c@{\hspace{2mm}}c}
			\hline
			\multicolumn{2}{c}{Trans.} & Wavelength & $\sigma$ & $\log gf$ & Exc. & W$_\lambda$ & $\log\epsilon_{^{12}\mathrm{C}}$ \\
			& & vac.~(nm) & (cm$^{-1}$) & & (eV) & (pm) & (3D) \\
			\hline
      P&32 & 2468.3280 & 4051.326 & $-$5.138 & 0.514 & 3.31 & 8.35 \\
      P&33 & 2471.9723 & 4045.353 & $-$5.126 & 0.530 & 3.30 & 8.36 \\
      P&36 & 2483.2111 & 4027.044 & $-$5.095 & 0.579 & 3.11 & 8.35 \\
      P&38 & 2490.9617 & 4014.514 & $-$5.075 & 0.614 & 3.01 & 8.35 \\
      P&40 & 2498.9218 & 4001.726 & $-$5.057 & 0.651 & 2.90 & 8.35 \\
      P&58 & 2580.5013 & 3875.216 & $-$4.928 & 1.065 & 1.51 & 8.35 \\
      P&60 & 2590.7333 & 3859.911 & $-$4.917 & 1.120 & 1.36 & 8.35 \\
      R&8  & 2359.4419 & 4238.290 & $-$5.590 & 0.283 & 2.14 & 8.35 \\
      R&13 & 2351.2311 & 4253.091 & $-$5.385 & 0.309 & 3.10 & 8.36 \\
      R&14 & 2349.7148 & 4255.836 & $-$5.353 & 0.315 & 3.26 & 8.36 \\
      R&15 & 2348.2403 & 4258.508 & $-$5.322 & 0.322 & 3.37 & 8.35 \\
      R&16 & 2346.8074 & 4261.108 & $-$5.293 & 0.330 & 3.50 & 8.35 \\
      R&17 & 2345.4162 & 4263.636 & $-$5.266 & 0.338 & 3.69 & 8.36 \\
      R&18 & 2344.0666 & 4266.090 & $-$5.234 & 0.346 & 3.79 & 8.36 \\
      R&19 & 2342.7584 & 4268.473 & $-$5.214 & 0.355 & 3.90 & 8.36 \\
      R&20 & 2341.4918 & 4270.782 & $-$5.190 & 0.365 & 4.03 & 8.36 \\
      R&22 & 2339.0825 & 4275.181 & $-$5.146 & 0.385 & 4.21 & 8.37 \\
      R&23 & 2337.9399 & 4277.270 & $-$5.124 & 0.396 & 4.25 & 8.36 \\
      R&24 & 2336.8385 & 4279.286 & $-$5.104 & 0.407 & 4.37 & 8.37 \\
      R&31 & 2330.2818 & 4291.327 & $-$4.975 & 0.499 & 4.89 & 8.41 \\
      R&32 & 2329.5096 & 4292.749 & $-$4.963 & 0.514 & 4.61 & 8.38 \\
      R&35 & 2327.4396 & 4296.567 & $-$4.914 & 0.562 & 4.50 & 8.37 \\
      R&37 & 2326.2652 & 4298.736 & $-$4.886 & 0.596 & 4.46 & 8.37 \\
      R&38 & 2325.7397 & 4299.707 & $-$4.870 & 0.614 & 4.54 & 8.39 \\
      R&40 & 2324.8122 & 4301.423 & $-$4.845 & 0.651 & 4.32 & 8.37 \\
      R&41 & 2324.4103 & 4302.167 & $-$4.830 & 0.670 & 4.28 & 8.37 \\
      R&56 & 2323.3645 & 4304.103 & $-$4.654 & 1.012 & 2.84 & 8.36 \\
      R&63 & 2326.1201 & 4299.004 & $-$4.583 & 1.206 & 2.19 & 8.37 \\
      R&64 & 2326.6855 & 4297.959 & $-$4.574 & 1.235 & 2.08 & 8.37 \\
      R&65 & 2327.2943 & 4296.835 & $-$4.564 & 1.265 & 2.14 & 8.40 \\
      R&68 & 2329.3824 & 4292.984 & $-$4.535 & 1.357 & 1.83 & 8.40 \\
      R&70 & 2330.9940 & 4290.015 & $-$4.517 & 1.421 & 1.58 & 8.38 \\
      R&77 & 2338.0411 & 4277.085 & $-$4.456 & 1.657 & 0.99 & 8.36 \\
      R&80 & 2341.7430 & 4270.324 & $-$4.431 & 1.765 & 0.79 & 8.35 \\
	\hline
	\end{tabular}
\end{minipage}%
\end{narrow}
\end{table*}

\bibliography{CObiblio}

\end{document}